\documentclass{aa}  
\usepackage{graphicx}
\usepackage[version=4]{mhchem}
\usepackage{units}
\usepackage{amsmath,amssymb}
\usepackage[utf8]{inputenc}
\usepackage[T1]{fontenc}
\usepackage{xcolor}
\usepackage{appendix}
\usepackage{booktabs}
\usepackage{multirow}
\DeclareUnicodeCharacter{2212}{-}
\usepackage{caption}
\usepackage[flushleft]{threeparttable}
\usepackage{graphicx}
\usepackage{txfonts}
\begin{document}

   \title{\ce{SO2} and \ce{OCS} toward high-mass protostars}

   \subtitle{A comparative study between ice and gas}

   \author{Julia C. Santos
          \inst{1}
          \and
          Martijn L. van Gelder\inst{2}
          \and
          Pooneh Nazari\inst{2,}\inst{3}
          \and
          Aida Ahmadi\inst{2}
          \and
          Ewine F. van Dishoeck\inst{2,}\inst{4}
          }

   %\date{Received September 15, 1996; accepted March 16, 1997}

    \institute{Laboratory for Astrophysics, Leiden Observatory, Leiden University, P.O. Box 9513, 2300 RA Leiden, NL\\
        \email{santos@strw.leidenuniv.nl}
    \and Leiden Observatory, Leiden University, P.O. Box 9513, 2300 RA Leiden, NL
    \and European Southern Observatory, Garching, DE
    \and Max-Planck-Institut f\"ur Extraterrestrische Physik, Giessenbachstrasse 1, D-85748 Garching, Germany
    }

   %\date{}

% \abstract{}{}{}{}{} 
% 5 {} token are mandatory
 
  \abstract
  % context heading (optional)
   {\ce{OCS} and \ce{SO2} are both major carriers of gaseous sulfur and are the only sulfurated molecules detected in interstellar ices to date. They are thus the ideal candidates to explore the evolution of the volatile sulfur content throughout the different stages of star formation.}
  % aims heading (mandatory)
   {We aim to investigate the chemical history of interstellar \ce{OCS} and \ce{SO2} by deriving a statistically-significant sample of gas-phase column densities towards massive protostars and comparing to observations of gas and ices towards other sources spanning from dark clouds to comets.}
  % methods heading (mandatory)
   {We analyze a subset of 26 line-rich massive protostars observed by ALMA in Band 6 as part of the High Mass Protocluster Formation in the Galaxy (ALMAGAL) survey. Column densities are derived for \ce{OCS} and \ce{SO2} from their rare isotopologues \ce{O^{13}CS} and \ce{^{34}SO2} towards the compact gas around the hot cores. We compare the abundance ratios of gaseous \ce{OCS}, \ce{SO2}, and \ce{CH3OH} with ice detections towards both high- and low-mass sources as well as dark clouds and comets.}
  % results heading (mandatory)
   {We find that gas-phase column density ratios of \ce{OCS} and \ce{SO2} with respect to methanol remain fairly constant as a function of luminosity between low- and high-mass sources, despite their very different physical conditions. In our dataset, \ce{OCS} and \ce{SO2} are weakly correlated. The derived gaseous \ce{OCS} and \ce{SO2} abundances relative to \ce{CH3OH} are overall similar to protostellar ice values, with a significantly larger scatter for \ce{SO2} than for \ce{OCS}. Cometary and dark-cloud ice values agree well with protostellar gas-phase ratios for \ce{OCS}, whereas higher abundances of \ce{SO2} are generally seen in comets compared to the other sources. Gaseous \ce{SO2}/\ce{OCS} ratios are consistent with ices toward dark clouds, protostars, and comets, albeit with some scatter.}
   % conclusions heading (optional), leave it empty if necessary 
   {The constant gas-phase column density ratios throughout low and high-mass sources indicate an early stage formation before intense environmental differentiation begins. Icy protostellar values are similar to the gas phase medians, compatible with an icy origin of these species followed by thermal sublimation. The larger spread in \ce{SO2} compared to \ce{OCS} ratios w.r.t. \ce{CH3OH} is likely due to a more water-rich chemical environment associated with the former, as opposed to a \ce{CO}-rich origin of the latter. Post-sublimation gas-phase processing of \ce{SO2} can also contribute to the large spread. Comparisons to ices in dark clouds and comets point to a significant inheritance of \ce{OCS} from earlier to later evolutionary stages.}
   
   \keywords{Astrochemistry, ISM: molecules, Stars: protostars, ISM: abundances, Techniques: interferometric}

   \maketitle

%%%%%%%%%%%%%%%%%%%%%%%%%%%%%%%%%%%%%%%%%%%%%%%%%%%%%%%%%%%%%%%%%%%%%%%%%%%%%%%%%%%%%%%%%%%%%%%%%%%%%%%%%%%%%%%%%%%%%%%%%%
%%%%%%%%%%%%%%%%%%%%%%%%%%%%%%%%%%%%%%%%%%%%%%%%%%%%%% INTRO %%%%%%%%%%%%%%%%%%%%%%%%%%%%%%%%%%%%%%%%%%%%%%%%%%%%%%%%%%%%%
%%%%%%%%%%%%%%%%%%%%%%%%%%%%%%%%%%%%%%%%%%%%%%%%%%%%%%%%%%%%%%%%%%%%%%%%%%%%%%%%%%%%%%%%%%%%%%%%%%%%%%%%%%%%%%%%%%%%%%%%%%
\section{Introduction}\label{sec:intro}
Over 240 molecules have been detected in the interstellar medium to date, among which at least 30 contain one or more sulfur atoms \citep{McGuire2022}. With an abundance of S/H$\sim$1.35$\times$10$^{−5}$, sulfur is one of the most common elements in space \citep{Asplund2009}. Indeed, S-bearing species are observed in the gas phase throughout most stages of star and planet formation, from diffuse and dense clouds (e.g., \citealt{Drdla1989, Navarro-Almaida2020, Spezzano2022, Esplugues2022}) and protostars (e.g., \citealt{Blake1987, Blake1994, vanderTak2003, Li2015, Drozdovskaya2018, Codella2021, delaVillarmois2023, Fontani2023, Kushwahaa2023}) to protoplanetary disks \citep{Fuente2010, Phuong2018, Semenov2018, LeGal2019, Riviere-Marichalar2021, LeGal2021, Booth2024}. They have also been detected in solar system bodies such as comets \citep{Smith1980, Bockelee-Morvan2000, Biver2021a, Biver2021b, Calmonte2016, Altwegg2022}, planets \citep{Moullet2013}, and satellites \citep{Hibbitts2000, Jessup2007, Moullet2008, Cartwright2020}, as well as towards extragalactic sources \citep{Henkel1985, Petuchowski1992, Mauersberger1995, Heikkila1999, Martin2003, Martin2005}. Identified species range from simple diatomic molecules such as \ce{CS} and \ce{SO} to the complex organics methanethiol (\ce{CH3SH}) and ethanethiol (\ce{\ce{CH3CH2SH}}) \citep{Linke1979, Gibb2000, Cernicharo2012, Koleniskova2014, Zapata2015, Muller2016, Majumdar2016, Rodriguez-Almeida2021}.

Despite this widespread detection, derived abundances in dense starless cores, protostars, and protoplanetary disks can only account for up to a few percent of the total expected cosmic value \citep{Tieftrunk1994, Wakelam2004, Anderson2013, Vastel2018, Fuente2019, LeGal2019, RiviereMarichalar2019, Riviere-Marichalar2020, LeGal2021, Bouscasse2022, Fuente2023}. The bulk of the sulfur content is largely thought to be locked away in or underneath the ice mantles that shroud interstellar dust grains, in a state that challenges its detection. Such icy mantles start to form early in the interstellar evolutionary sequence, during the so-called translucent-cloud phase. Atoms of \ce{H} and \ce{O} adsorb onto dust grains and react to form \ce{H2O}, resulting in a water-rich ice layer \citep{Tielens1982, Hiraoka1998, Mokrane2009, Dulieu2010, Ioppolo2010, Cuppen2010, Romanzin2011, Oberg2011}. As the density of the collapsing cloud increases, carbon monoxide (\ce{CO}) molecules present in the gas phase catastrophically freeze-out on top of the water-rich ice, forming a second coating known as the CO-rich ice layer \citep{Tielens1991, Boogert2002, Pontoppidan2003, Pontoppidan2006, Oberg2011}. This \ce{CO} ice is efficiently converted into methanol (\ce{CH3OH}) by reactions involving \ce{H} atoms \citep{Tielens1982, Charnley1992, Hiraoka1994, Watanabe2002, Fuchs2009, Cuppen2009, Santos2022}. Complementarily, some smaller contribution to \ce{CH3OH} formation from ice chemistry in less dense environments, before the catastrophic CO freeze-out, is also often invoked (e.g., \citealt{Wada2006, Hodyss2009, Oberg2010, Bergner2017, Lamberts2017, Qasim2018}), although it is likely not dominant. 

Eventually, the environment close to the emerging young stellar object is warmed up to temperatures of $100-300$ K, resulting in the complete thermal sublimation of the ices. This chemically-rich region surrounding the protostar is known as hot core for massive sources or hot corino for low-mass counterparts (e.g., \citealt{Herbst2009}), and is thought to be representative of the bulk ice content. For simplicity, we henceforth utilize ``hot core'' as an umbrella term for both high- and low-mass sources. By studying the compact emission originated from the hot core in comparison to ice observations, it is possible to gain insight into the formation and destruction mechanisms of molecules both in the gas and in the solid phases.

Compared to gaseous species, the unambiguous detection of solid-state molecules embedded in interstellar ices poses significantly more challenges. Spectral features of species in the solid phase are intrinsically broad and highly degenerate, with properties such as peak position and width that vary considerably with the ice environment. Likely as a consequence of such inherent limitations, only two sulfur-bearing species have been identified in ices so far: carbonyl sulfide (\ce{OCS}) and sulfur dioxide (\ce{SO2}). The former was first detected by \cite{Palumbo1995} towards the massive protostar W33A. Soon after, \cite{Boogert1997} suggested the presence of the latter towards both W33A and NGC 7538:IRS1---another massive young stellar object (MYSO). Since then, both species have been either inferred or detected in ices towards other protostars as well as dark clouds \citep{Palumbo1997, Oberg2008, Zasowski2009, Yang2022, Boogert2022, McClure2023, Rocha2024}. Recently, the presence of \ce{SO2} ice was confirmed by JWST observations towards the solar-type protostar IRAS 2A by constraining the contribution of blended species, in particular \ce{OCN^-}, to its 7.6 $\mu$m region \citep{Rocha2024}.

In the gas phase, both species are commonly detected. Gaseous \ce{SO2} is observed towards protostellar systems both through its pure rotational transitions occurring at submillimeter wavelengths and through its rovibrational lines probed by the mid-infrared \citep{Keane2001, Dungee2018, Nickerson2023, vanGelder2024}. It is a good tracer of outflows, jets, and accretion shocks due to its enhanced gas-phase formation at high temperatures ($T\gtrsim$100 K) combined with either sputtering or thermal sublimation  of \ce{SO2} or its precursors from icy dust grains \citep{Pineau1993, Sakai2014, Oya2019, Taquet2020, Tychoniec2021, vanGelder2021}. It has also been shown to trace disk winds \citep{Tabone2017}. While this is the case for the main isotopologue (i.e., \ce{^{32}SO2}), the emission of minor isotopologues such as \ce{^{34}SO2} is expected to be mostly compact, tracing the hot core region. Moreover, in contrast to \ce{SO2}, gaseous \ce{OCS} emission is not expected to have significant contributions from outflows \citep{vanderTak2003, Drozdovskaya2018}, tracing predominately the envelope surrounding protostars (e.g., \citealt{Herpin2009, Oya2016}). Being the only two sulfurated molecules detected in both gas and ices, \ce{SO2} and \ce{OCS} are the ideal targets to perform a comparative study between these two physical states.

In this work we explore the origin and fate of two of the most abundant sulfur species, \ce{OCS} and \ce{SO2}, by directly comparing their solid and gaseous components during the evolution of star forming regions. We utilize data from the ALMA Evolutionary study of High Mass Protocluster Formation in the Galaxy (ALMAGAL) survey and select a subsample of 26 line-rich sources to perform the analysis, representing the first interferometric study on a statistically significant sample of detections for these two molecules in the gas. The molecular column densities are derived from the rare isotopologues \ce{O^{13}CS} and \ce{^{34}SO2} to avoid contamination from extended emission, outflows, and to limit line optical depth effects. We thus focus on the hot core region, which contains the molecular reservoir from the ices after complete thermal sublimation. Their ratios are compared to other observations in the solid and the gas phases taken from the literature. This includes recent works that utilize ground-based infrared observatories to investigate \ce{OCS} ice abundances (among other species) in a large sample of MYSOs \citep{Boogert2022}, as well as space observations by the \textit{James Webb} Space Telescope (JWST) of both \ce{OCS} and \ce{SO2} ices towards sources ranging from background stars to protostars \citep{McClure2023, Rocha2024}. This study sets the stage for future gas and ice works enabled by the combination of data from the Atacama Large Millimeter/submillimeter Array (ALMA) and JWST on both low- and high-mass protostars, paving the way to a more thorough understanding of the sulfur chemical evolution of the interstellar medium.

In Section \ref{sec:observations} we describe the observational parameters and details of the dataset, as well as the procedure to fit a synthetic spectrum to the lines of interest. The resulting emission morphologies, kinematics, and column densities are presented in Section \ref{sec:results}. The column density ratios with respect to methanol are discussed in Section \ref{sec:discussion} in comparison to other observations. Finally, our main findings are summarized and concluded in Section \ref{sec:conc}.
%%%%%%%%%%%%%%%%%%%%%%%%%%%%%%%%%%%%%%%%%%%%%%%%%%%%%%%%%%%%%%%%%%%%%%%%%%%%%%%%%%%%%%%%%%%%%%%%%%%%%%%%%%%%%%%%%%%%%%%%%%
%%%%%%%%%%%%%%%%%%%%%%%%%%%%%%%%%%%%%%%%%%%%%%%%%%%%% METHODS %%%%%%%%%%%%%%%%%%%%%%%%%%%%%%%%%%%%%%%%%%%%%%%%%%%%%%%%%%%%
%%%%%%%%%%%%%%%%%%%%%%%%%%%%%%%%%%%%%%%%%%%%%%%%%%%%%%%%%%%%%%%%%%%%%%%%%%%%%%%%%%%%%%%%%%%%%%%%%%%%%%%%%%%%%%%%%%%%%%%%%%
\section{Observations and methods}\label{sec:observations}
\subsection{The observations}\label{subsec:data}
The ALMAGAL survey (2019.1.00195.L; PIs: P. Schilke, S. Molinari, C. Battersby, P. Ho) was observed by ALMA in Band 6 ($\sim$1 mm). It targeted over 1000 dense clumps across the Galaxy with $M>500$ M$_\odot$ and $d<7.5$ kpc, chosen based on the Herschel infrared Galactic Plane Survey (Hi-GAL) \citep{Molinari2010, Elia2017, Elia2021}. ALMAGAL covers a statistically significant sample of sources in all stages of star formation, with many of them consisting of MYSOs. For this work we consider only archival data made publicly available before February 2021, with beam sizes between 0.5\arcsec and 1.5\arcsec ($\sim$1000–5000 au). The Common Astronomy
Software Applications\footnote{https://casa.nrao.edu} (CASA; \citealt{McMullin2007}) version 5.6.1. was used to pipeline calibrate and image the data. The selection of sources was based on a subset studied in both \cite{Nazari2022} and \cite{vanGelder2022}, thus culling for particularly line-rich sources for which the \ce{CH3OH} column densities---our benchmark for comparison---are well constrained. In \cite{vanGelder2022}, the selection was based on sources with high bolometric luminosities ($L_{\text{bol}}>1000$ $L_\odot$) where complex organic molecules (COMs) such as \ce{CH3OH} and \ce{CH3CN} were detected. In \cite{Nazari2022}, the selection criterium consisted of sources that contain the \ce{CH3CN} $12_7-11_7$ line above the $2.5-3\sigma$ level. Within this subset commonly pertaining to both studies, we exclude sources for which the line profiles of the targeted molecules differed significantly across species, to avoid probing distinct emitting regions. An example case of an excluded source is shown in Appendix \ref{appendix:bad_source}. The final subset consists of 26 line-rich sources, whose properties and observational parameters are listed in Appendix \ref{appendix:source_stuff}.

In this work we make use of two out of the four spectral windows in ALMAGAL, encompassing frequencies of $\sim$217.00$-$218.87 GHz and $\sim$219.07$-$220.95 GHz with a spectral resolution of $\sim$0.5 MHz ($\sim$0.7 km s$^{-1}$). The spectral windows cover 4 transitions of \ce{^{34}SO2}, 11 transitions of \ce{^{33}SO2}, and 1 transition of \ce{O^{13}CS} with upper energy $E_{\text{up}}<800$ K and Einstein $A_{\text{ij}}>1\times10^{-6}$ s$^{-1}$ (see Appendix \ref{appendix:transitions}). However, only the $11_{1,11}-10_{0,10}$ transitions of \ce{^{34}SO2} and \ce{^{33}SO2} were detected in our sources (E$_{\text{up}}$ = 60.1 K and 57.9 K, respectively). For \ce{^{34}SO2}, this corresponds to only one line, whereas for \ce{^{33}SO2} it encompasses 7 hyperfine components caused by the nuclear spin $I=3/2$ of the \ce{^{33}S} atom. The remaining \ce{^{34}SO2} and \ce{^{33}SO2} lines are rather weak ($A_{ij}<2.6\times10^{-5}$ s$^{-1}$) and for the most part highly blended, hampering their detections. \ce{O^{13}CS} is detected in its $18-17$ transition (E$_{\text{up}}$ = 99.5 K). No other isotopologues of \ce{OCS}, including \ce{O^{12}CS}, are covered in the ALMAGAL range. One line of \ce{^{32}SO2} is covered, but it is not included in this work due to its likely contamination from outflow emission and because it is probably optically thick.

The spectra utilized in this work are the same as in \cite{vanGelder2022}. For all sources with the \ce{CH3OH} $8_{0,8}-7_{1,6}$ line ($E_{\text{up}}=97$ K) above the $3\sigma$ level, the spectra were extracted from the peak pixel for this line in the integrated intensity maps. This particular transition of \ce{CH3OH} was chosen because it is the strongest methanol line within the sample with $E_{\text{up}}>70$ K, to avoid contamination by the outflow or extended emission that are probed by lines with lower $E_{\text{up}}$. The choice of extracting the spectra from the methanol peak intends to maximize the signal-to-noise ratio of both \ce{^{34}SO2} and \ce{O^{13}CS} originating from the hot core region. In G023.3891p00.1851, the peak emission of the methanol isotopologue \ce{CH2DOH} is offset by $\sim0.6\arcsec$ (approximately half a beam) from that of \ce{CH3OH}, and thus the \ce{CH2DOH} $17_{1,16}e_0-17_{0,17}e_0$ ($E_{\text{up}}=336$ K) peak is chosen to extract the spectrum since it is a more reliable tracer of the hot core. All sources probed here show line widths of $\gtrsim3$ km s$^{-1}$, well above the spectral resolution of $\sim0.7$ km s$^{-1}$, and the spectral sensitivity corresponds to $\sim0.2$ K (for a list of rms values see Appendix \ref{appendix:source_stuff}). 

%%%%%%%%%%%%%%%%%%%%%%%%%%%%%%%%%%%%%%%%%%%%%%%%%%%%%%%%%%%%%%%%%%%%%%%%%%%%%%%%%%%%%%%%%%%%%%%%%%%%%%%%%%%%%%%%%%%%%%%%%%
\subsection{Spectral analysis}\label{subsec:spectral_analysis}
The target molecules of this study are \ce{SO2} and \ce{OCS}, the two sulfur-bearing species detected in both gas and ices so far. We analyze them by means of their rare isotopologues \ce{^{34}SO2} and \ce{O^{13}CS} in order to avoid issues with optically-thick lines and contamination from extended emission and outflows. The \ce{^{33}SO2} emission is also analyzed as a diagnostic tool to assess whether or not \ce{^{34}SO2} is indeed optically thin (Section \ref{subsec:34SO2_opt_thick}). We utilize the CASSIS\footnote{http://cassis.irap.omp.eu/} spectral analysis tool \citep{Vastel2015} to fit the spectrum for each source and derive each species' column density ($N$) and full width at half maximum (FWHM) assuming that the excitation is under local thermodynamic equilibrium (LTE). Since we only detect one line per species per source, the excitation temperature is fixed to $T_{\text{ex}}=150$ K in the spectral fittings---a roughly averaged value for hot cores (see, e.g., \citealt{vanGelder2020, Yang2021, Nazari2022}). Fixing the $T_{\text{ex}}$ to $60-250$ K only changes the derived column densities by up to a factor of $\sim$ 3, and both molecules in the same direction. Furthermore, for $\gtrsim80\%$ of the sources in this work, excitation temperatures derived from \ce{CH3CN} lines range between $120 - 170$ K \citep{Nazari2022}, corresponding to only up to a factor of 1.6 difference in column densities for \ce{SO2} and \ce{OCS}. Thus, the assumption of a fixed $T_{\text{ex}}=150$ K does not interfere significantly with the column density ratios---the main focus of this work. The spectroscopic properties employed in the fittings of each species are obtained from the Cologne Database for Molecular Spectroscopy (CDMS; \citealt{Muller2001, Muller2005}).

The \ce{^{34}SO2} and \ce{O^{13}CS} lines analyzed in this work are mostly unblended, which allows the utilization of the grid fitting method as explained in detail previously \citep{vanGelder2020, Nazari2021, Chen2023}. In summary, a grid of column densities and FWHMs is tested and the best fit model is assigned to the combination with the lowest $\chi ^2$. In this work $N$ was varied from $1\times10^{13}$ cm$^{−2}$ to $1\times10^{17}$ cm$^{−2}$ with a step of 0.1 on a logarithmic scale, and the FWHM was varied from 3 km s$^{−1}$ to 11 km s$^{−1}$ with a spacing of 0.1 km s$^{−1}$ on a linear scale. The radial velocities ($V_{\text{lsr}}$) are derived by eye using increments of 0.1 km s$^{-1}$ in a similar manner as described in \cite{Nazari2022} and \cite{Chen2023}, and are fixed to the best manually-derived values for the grid fits. Their median offsets are of 0.35 km s$^{-1}$ from the velocities for \ce{CH3OH} in the same sources (see Section \ref{subsec:fitting_results}). The $2\sigma$ errors are derived from the reduced $\chi ^2$ calculated from the comparison between the resulting model of each grid point and the observed spectrum. If the $2\sigma$ uncertainties of the column densities are smaller than 20\%, we assume a 20\% uncertainty as a conservative estimate to account for systematic sources of errors. In some few instances, such as for \ce{^{34}SO2} in 693050, more severe blending or deviations from a Gaussian profile are observed (see Appendix \ref{appendix:best_models}). Nonetheless, the integrated line emissions are still encompassed by our models well within the adopted conservative uncertainties, and thus it does not affect our analysis. In the cases of G025.6498p01.0491 for \ce{^{34}SO2}, and 126348 and 707948 for \ce{O^{13}CS}, the emission is better described by two components (see Appendix \ref{appendix:best_models}). However, obtaining column density ratios for each velocity component separately is not possible since they are not resolved for methanol. In such cases, we perform the grid fits assuming one Gaussian and fixing the $V_{\text{lsr}}$ to the mean between the values derived by eye for each component. This procedure yields column densities within 20$\%$ of the ones obtained by manually fitting each component separately. Thus, to ensure a systematic approach to the line analysis, we utilize the column densities derived from the grid fittings to these sources. The exact source sizes are not known, so we assume that the source fully fills the beam (i.e., a beam dilution factor of unity). Since this work focuses on comparing column density ratios, this assumption does not interfere with the analysis as long as the lines are optically thin \citep{vanGelder2020, Nazari2021}.

Fitting \ce{^{33}SO2} is more challenging. It is considerably less abundant than \ce{^{34}SO2}, and its only detected transitions are blended with \ce{CH3^{13}CN}. The fitting was thus performed by eye on top of the best models for \ce{CH3^{13}CN} derived by \cite{Nazari2022}. The $V_{\text{lsr}}$ values of \ce{^{33}SO2} were fixed to those of \ce{^{34}SO2}, while its column densities and FWHMs were varied in steps of 0.1 in log space and 0.1 in linear space, respectively. For most sources, barely or no emission was left underfitted after accounting for \ce{CH3^{13}CN}. In such cases we assign \ce{^{33}SO2} as upper limits. However, for four sources (615590, 644284A, 693050, G343.1261-00.0623) the \ce{CH3^{13}CN} model underfitted the emission significantly ($\lesssim$50\%), in which cases we can derive approximate column densities for \ce{^{33}SO2}. 

\begin{figure*}[htb!]\centering
\includegraphics[scale=0.7]{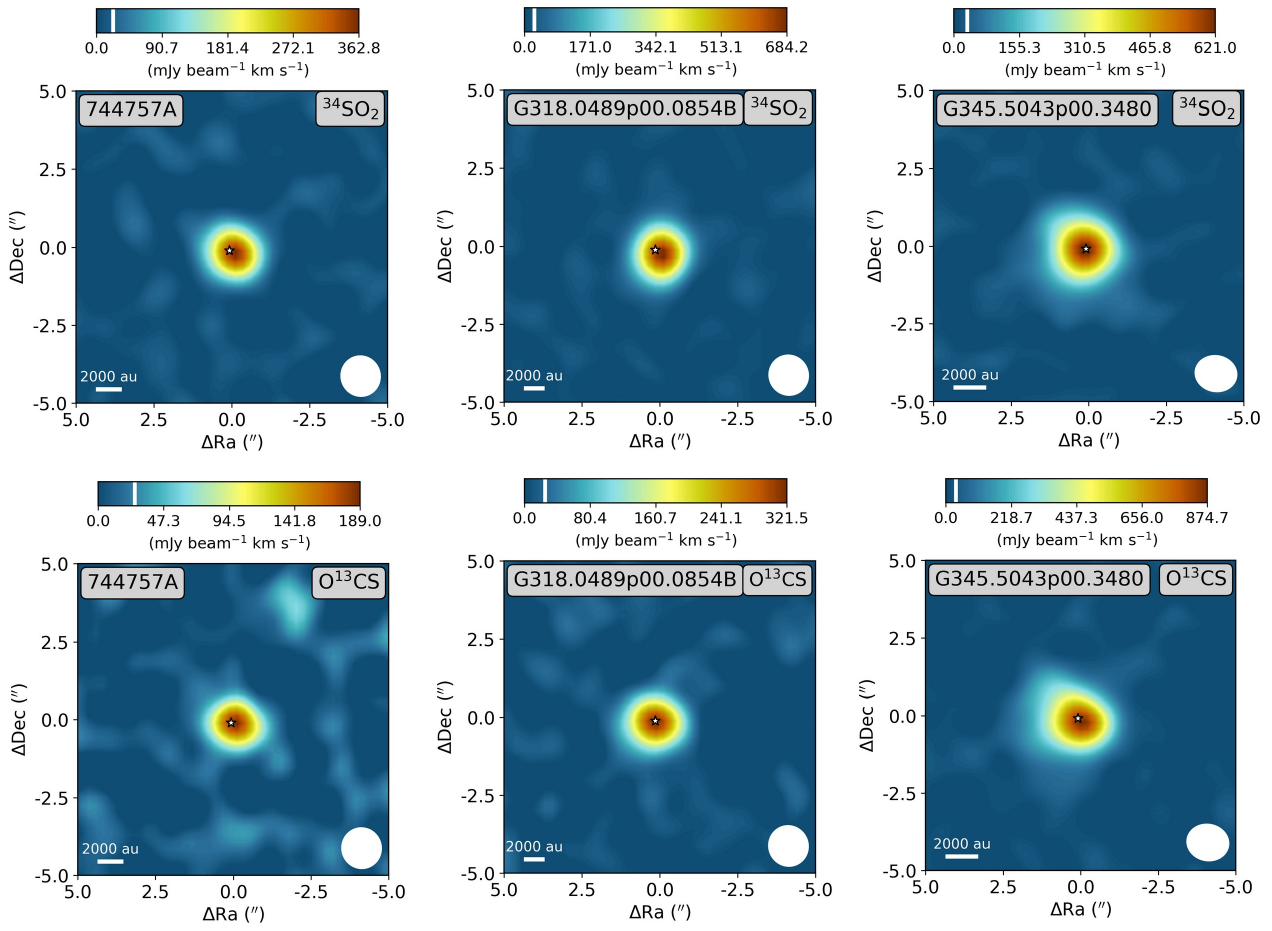}
\caption{Integrated intensity maps of the \ce{^{34}SO2} 11$_{1,11}$ $-$ 10$_{0,10}$ (E$_{\text{up}}$ = 60.1 K, top) and \ce{O^{13}CS} 18 $-$ 17 (E$_{\text{up}}$ = 99.5 K, bottom) lines for 744757A, G318.0489p00.0854B, and G345.5043p00.3480. The integration limits are set to [-2, 2] km s$^{-1}$ with respect to the sources' $V_{\text{lsr}}$. The white star denotes the source positions derived from the peak continuum emission and the $3\sigma$ threshold is denoted by the white line in the color bars. The beam size is shown in the lower right corner of each panel, and a scale bar is depicted in the lower left.}
\label{fig:moment0_subset}
\end{figure*}
%%%%%%%%%%%%%%%%%%%%%%%%%%%%%%%%%%%%%%%%%%%%%%%%%%%%%%%%%%%%%%%%%%%%%%%%%%%%%%%%%%%%%%%%%%%%%%%%%%%%%%%%%%%%%%%%%%%%%%%%%%
\subsection{Isotope ratio callibration}\label{subsec:isotope_ratios}
Isotopic abundances are dependent on the stellar population and therefore vary as a function of the distance to the Galactic center ($D_{\text{GC}}$). In order to obtain accurate column densities of the main \ce{SO2} and \ce{OCS} isotopologues, it is thus required to calibrate the isotope ratios of (\ce{^{32}S}/\ce{^{34}S}), (\ce{^{32}S}/\ce{^{33}S}), and (\ce{^{12}C}/\ce{^{13}C}) accordingly. Recently, \cite{Yan2023} utilized \ce{CS} lines in a wide variety of isotopologues observed towards 110 high-mass star forming regions to derive the equations:
\begin{equation}
(^{32}\text{S}/^{34}\text{S}) = (0.73 \pm 0.36)D_\text{GC} + (16.50 \pm 2.07),
\label{eq:32S_34S}
\end{equation}

\begin{equation}
(^{32}\text{S}/^{33}\text{S}) = (2.64 \pm 0.77)D_\text{GC} + (70.80 \pm 5.57),
\label{eq:32S_33S}
\end{equation}

\begin{equation}
(^{12}\text{C}/^{13}\text{C}) = (4.77 \pm 0.81)D_\text{GC} + (20.76 \pm 4.61).
\label{eq:12C_13C}
\end{equation}

$D_{\text{GC}}$ can be calculated based on the source's distance to Earth ($d$) and its coordinates. The resulting values for each source are listed in Appendix \ref{appendix:source_stuff} and are used to derive \ce{^{32}SO2} and \ce{O^{12}CS} from the isotopologues. For the solar neighborhood, the ratios are (\ce{^{32}S}/\ce{^{34}S})$\sim$22, (\ce{^{32}S}/\ce{^{33}S})$\sim$92, and (\ce{^{12}C}/\ce{^{13}C})$\sim$59. The uncertainties in the final column densities of \ce{^{32}SO2} and \ce{O^{12}CS} (henceforth simply \ce{SO2} and \ce{OCS}) are obtained by propagating the errors in the derived column densities of the minor isotopologues together with the uncertainties in Equations \ref{eq:32S_34S}$-$\ref{eq:12C_13C} and the typical error of $\sim$0.5 kpc in D$_\text{GC}$ (see \citealt{Nazari2022}).
%%%%%%%%%%%%%%%%%%%%%%%%%%%%%%%%%%%%%%%%%%%%%%%%%%%%%%%%%%%%%%%%%%%%%%%%%%%%%%%%%%%%%%%%%%%%%%%%%%%%%%%%%%%%%%%%%%%%%%%%%%
%%%%%%%%%%%%%%%%%%%%%%%%%%%%%%%%%%%%%%%%%%%%%%%%%%%%% RESULTS %%%%%%%%%%%%%%%%%%%%%%%%%%%%%%%%%%%%%%%%%%%%%%%%%%%%%%%%%%%%
%%%%%%%%%%%%%%%%%%%%%%%%%%%%%%%%%%%%%%%%%%%%%%%%%%%%%%%%%%%%%%%%%%%%%%%%%%%%%%%%%%%%%%%%%%%%%%%%%%%%%%%%%%%%%%%%%%%%%%%%%%
\section{Results}\label{sec:results}
\vspace{-5px}
\subsection{Morphology}\label{subsec:morph}
The integrated intensity maps of \ce{^{34}SO2} and \ce{O^{13}CS} for 744757A, G318.0489p00.0854B, and G345.5043p00.3480 are presented in Figure \ref{fig:moment0_subset}. These are chosen as a representative sample of the sources analyzed in this work. The emission areas of both species are compact (with typical radii of $\sim1000-3000$ au considering all sources) and mostly unresolved, in accordance with the expectation that rare isotopologues likely trace the hot core region with little to no contribution from extended emission. The case of 693050 is an exception in which the \ce{^{34}SO2} and \ce{O^{13}CS} peaks are offset from the continuum peak (as shown by the white star) by $\sim$4000 au (see Appendix \ref{appendix:moment0_693050}). This could be the result of optically thick dust at these wavelengths, leading to continuum oversubstraction or dust attenuation towards the continuum peak, which has been shown to hide molecular emission in prostellar systems \citep{deSimone2020}. Indeed, the same behavior is observed for \ce{CH3OH} emission in 693050, in line with this hypothesis \citep{vanGelder2022a}. Higher spatial resolution is required to fully distinguish the species' emitting regions. Still, some spatial information can be acquired by comparing their best-fit parameters (see Section \ref{subsec:fitting_results}).
%%%%%%%%%%%%%%%%%%%%%%%%%%%%%%%%%%%%%%%%%%%%%%%%%%%%%%%%%%%%%%%%%%%%%%%%%%%%%%%%%%%%%%%%%%%%%%%%%%%%%%%%%%%%%%%%%%%%%%%%%%
\begin{figure*}[htb!]\centering
\includegraphics[scale=0.4]{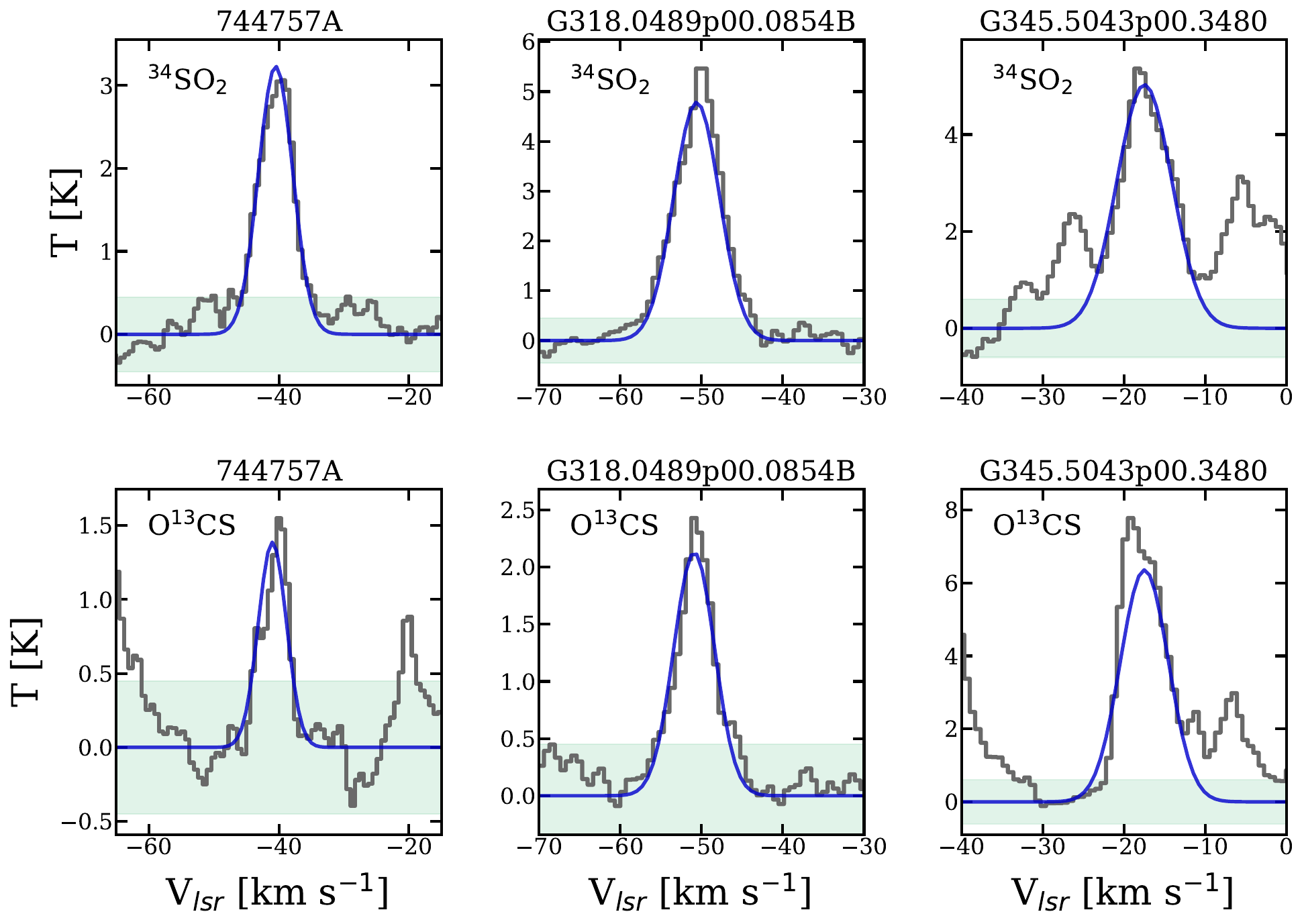}
\caption{Observed spectra towards 744757A, G318.0489p00.0854B, and G345.5043p00.3480 (grey) superimposed by their best-fit models (blue). The upper panels show the lines of \ce{^{34}SO2} ($11_{1,11} - 10_{0,10}$), and the lower panels lines of \ce{O^{13}CS} ($18-17$). The green shadowed area delimits the $3\sigma$ threshold.}
\label{fig:fittings_subset}
\end{figure*}

\subsection{Fitting results}\label{subsec:fitting_results}
The best-fit parameters of \ce{^{34}SO2} and \ce{O^{13}CS} for all sources are listed in Appendix \ref{appendix:fits}, together with the corresponding column densities derived for \ce{SO2} and \ce{OCS}. The best-fit models are shown in Figure \ref{fig:fittings_subset} for a set of representative sources: 744757A, G318.0489p00.0854B, and G345.5043p00.3480. The models for the remaining sources are presented in Appendix \ref{appendix:best_models}. $3\sigma$ upper limits are provided when line intensities do not surpass this detection threshold. For most sources, the \ce{^{34}SO2} column densities range between $10^{15}-10^{16}$ cm$^{-2}$, which corresponds to \ce{SO2} column densities of $10^{16}-10^{17}$ cm$^{-2}$. The only exceptions to this trend are the sources for which only $N$(\ce{^{34}SO2}) upper limits could be derived.

In comparison, the \ce{O^{13}CS} column densities are roughly one order of magnitude lower, but given the larger ratios of (\ce{^{12}C}/\ce{^{13}C}) compared to (\ce{^{32}S}/\ce{^{34}S}), the corresponding $N$(\ce{OCS}) values also range between $10^{16}-10^{17}$ cm$^{-2}$. Absolute column densities are nevertheless subject to biases such as the assumed emitting area and thus are not ideal to be directly compared. Column density ratios are a more reliable form of comparison to provide information on the chemical inventories of different systems. The column densities derived in this work will be discussed in detail in Section \ref{sec:discussion}.

Despite the emissions probed here being largely unresolved, line widths and velocities can be utilized to infer the kinematics of the gas. In Appendix \ref{appendix:FWHMs_Vlsrs_wrtCH3OH}, we present a comparison between the widths and velocities for both \ce{^{34}SO2} and \ce{O^{13}CS} with respect to \ce{CH3^{18}OH}. The latter was obtained from the fittings performed by \cite{vanGelder2022}. For some sources, \cite{vanGelder2022} did not detect \ce{CH3^{18}OH}, in which cases we compare to the fitting parameters of the main \ce{CH3OH} isotopologue (signaled by empty markers). The median offsets of the line widths and peak velocities between the three different species are, respectively, 0.15 and 0.35 km s$^{-1}$. This is in line with all three molecules tracing a similar compact gas within the hot core region and serves as validation for a comparative analysis of their column density ratios. The fact that no distinctively large discrepancy in FWHM and $V_{\text{lsr}}$ is observed for \ce{^{34}SO2} is an indication that any contribution from outflows to this line can be neglected.
%%%%%%%%%%%%%%%%%%%%%%%%%%%%%%%%%%%%%%%%%%%%%%%%%%%%%%%%%%%%%%%%%%%%%%%%%%%%%%%%%%%%%%%%%%%%%%%%%%%%%%%%%%%%%%%%%%%%%%%%%%
\subsection{Is \ce{^{34}SO2} optically thin?}\label{subsec:34SO2_opt_thick} 
\begin{figure}[htb!]\centering
\includegraphics[scale=0.26]{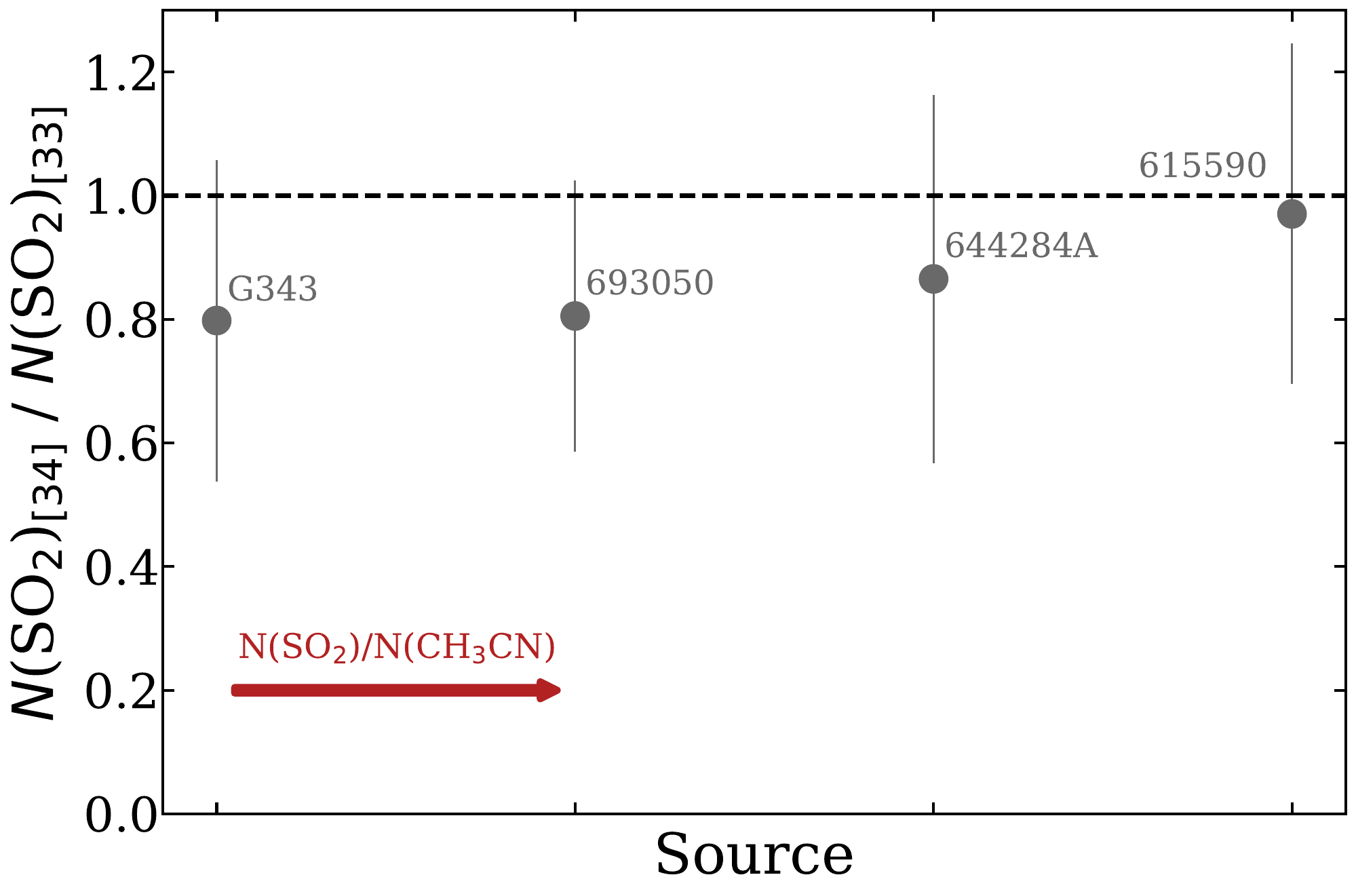}
\caption{Ratios of $N$(\ce{^{32}SO2}) derived from $N$(\ce{^{34}SO2}) (numerator) over those derived from $N$(\ce{^{33}SO2}) (denominator). The red arrow indicates that the sources are sorted in order of increasing $N$(\ce{SO2})/$N$(\ce{CH3CN}) relative abundances The dashed line highlights the unity mark. G343 stands for source G343.1261-00.0623.}
\label{fig:SO2_34_33}
\end{figure}

Line emissions from rare isotopologues are usually assumed to be optically thin. However, this is not necessarily always true, especially for abundant species such as \ce{SO2}. For this reason, we utilize the four sources in which \ce{^{33}SO2} is detected to assess whether \ce{^{34}SO2} is indeed optically thin. Their best-fit parameters are listed in Appendix \ref{appendix:fitting_33SO2}. We derive the column densities of \ce{SO2} (i.e., of the main isotopologue) from both $N$(\ce{^{34}SO2}) and $N$(\ce{^{33}SO2}) separately, using their isotope ratios as described in Section \ref{subsec:isotope_ratios}. Figure \ref{fig:SO2_34_33} shows the ratios of \ce{SO2} column densities derived from \ce{^{34}SO2} over the \ce{^{33}SO2} counterparts for the four sources. The resulting values are all remarkably close to unity, confirming that both the \ce{^{34}SO2} and \ce{^{33}SO2} lines are indeed optically thin. In fact, the larger the $N$(\ce{SO2})/$N$(\ce{CH3CN}) ratios (indicated by the red arrow), the closer the values are to unity. Given that the \ce{^{33}SO2} lines had to be fitted on top of \ce{CH3^{13}CN} (see the discussion in Section \ref{subsec:spectral_analysis}), this strongly suggests that the small discrepancies in \ce{SO2} column densities calculated from $N$(\ce{^{33}SO2}) and $N$(\ce{^{34}SO2}) are mostly due to \ce{^{33}SO2} being heavily blended.

%%%%%%%%%%%%%%%%%%%%%%%%%%%%%%%%%%%%%%%%%%%%%%%%%%%%%%%%%%%%%%%%%%%%%%%%%%%%%%%%%%%%%%%%%%%%%%%%%%%%%%%%%%%%%%%%%%%%%%%%%%
%%%%%%%%%%%%%%%%%%%%%%%%%%%%%%%%%%%%%%%%%%%%%%%%%%%% DISCUSSION %%%%%%%%%%%%%%%%%%%%%%%%%%%%%%%%%%%%%%%%%%%%%%%%%%%%%%%%%%
%%%%%%%%%%%%%%%%%%%%%%%%%%%%%%%%%%%%%%%%%%%%%%%%%%%%%%%%%%%%%%%%%%%%%%%%%%%%%%%%%%%%%%%%%%%%%%%%%%%%%%%%%%%%%%%%%%%%%%%%%%
\section{Discussion}\label{sec:discussion}
As mentioned in Section \ref{subsec:fitting_results}, column density ratios are a good metric to compare the chemical content of different sources and types of environments. Here we utilize the \ce{CH3OH} column densities derived by \cite{vanGelder2022} from minor isotopologues as a basis for comparison. Given the ice origin of methanol, it is possible to deduct information on the chemical history of \ce{OCS} and \ce{SO2} from their relative abundances with respect to \ce{CH3OH}. In the following subsections, we compare $N$(\ce{OCS})/$N$(\ce{CH3OH}) and $N$(\ce{SO2})/$N$(\ce{CH3OH}) derived in this work with relative abundances taken from the literature. These encompass gas-phase observations towards both massive and low-mass young stellar objects (MYSOs and LYSOs), ice observations towards protostars and dark clouds, as well as cometary ratios. A complete list of references from which these ratios are taken can be found in Appendix \ref{appendix:lit_ratios}.
%%%%%%%%%%%%%%%%%%%%%%%%%%%%%%%%%%%%%%%%%%%%%%%%%%%%%%%%%%%%%%%%%%%%%%%%%%%%%%%%%%%%%%%%%%%%%%%%%%%%%%%%%%%%%%%%%%%%%%%%%%
\subsection{\textit{N}(\ce{OCS})/\textit{N}(\ce{CH3OH})}\label{subsec:NOCS_NCH3OH}
\begin{figure*}[htb!]\centering
\includegraphics[scale=0.42]{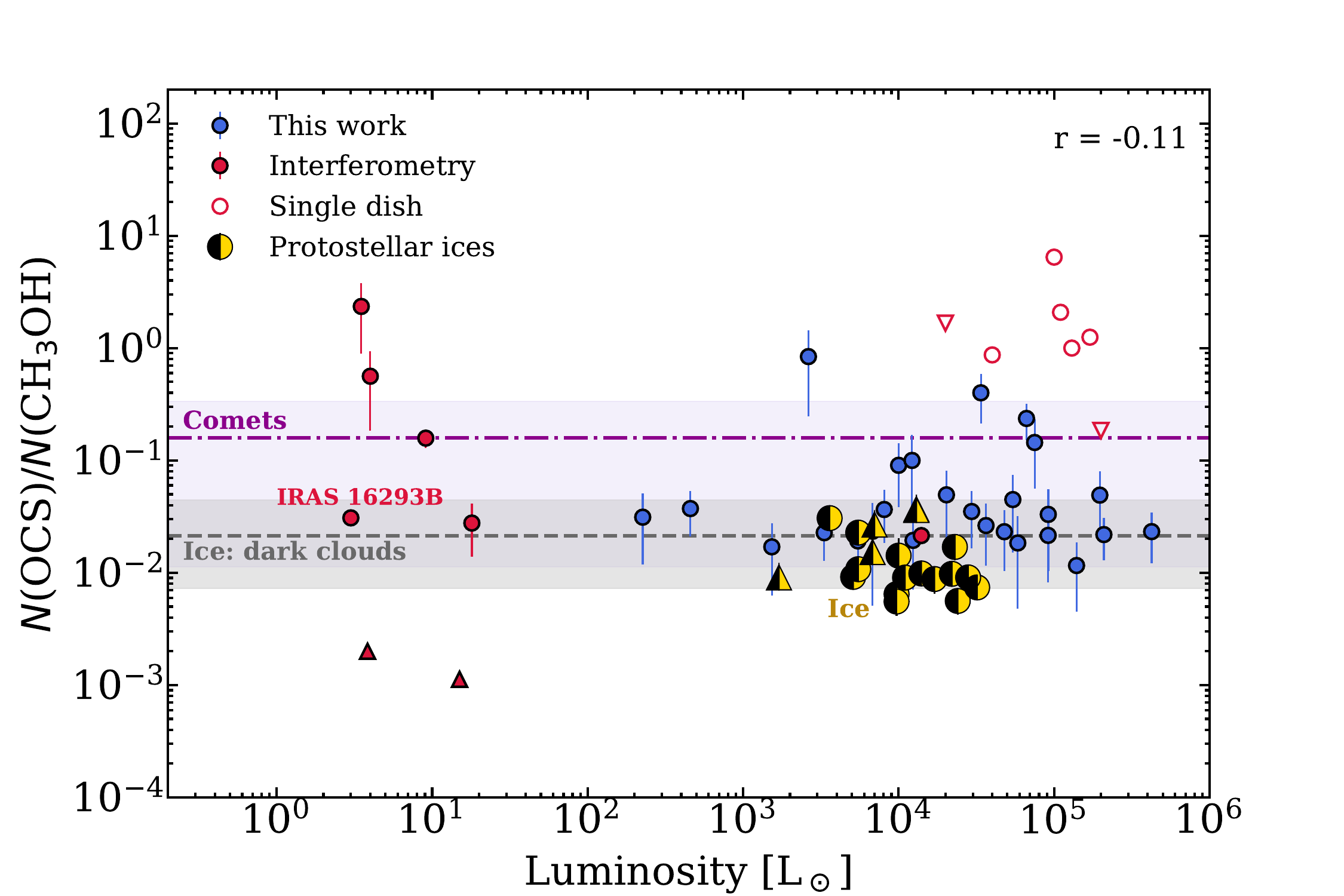}
\caption{Column density ratios $N$(\ce{OCS})/$N$(\ce{CH3OH}) derived in this work (blue markers) for 26 high-mass protostars as a function of luminosity. Literature gas-phase values are shown for comparison (red markers), together with ice values in protostars (yellow and black markers), dark clouds (gray dashed line), and comets (purple dash-dot line). The references can be found in Appendix \ref{appendix:lit_ratios}. Upper and lower limits towards protostars are denoted by downward- and upward-facing triangles, respectively. The range of values for dark clouds and comets are shown by their respective shadowed areas. For gas-phase ratios, filled markers correspond to interferometric observations, whereas empty markers denote single dish counterparts. The Pearson correlation coefficient for protostellar ratios in both gas and ices, but excluding single-dish observations and lower limits, is displayed in the upper right corner. IRAS 16293B stands for the source IRAS 16293-2422 B.}
\label{fig:NOCS_NCH3OH}
\end{figure*}

%  \begin{figure*}
% \sidecaption
%   \includegraphics[width=12cm]{Figures/NOCS_NCH3OH.pdf}
%      \caption{Column density ratios $N$(\ce{OCS})/$N$(\ce{CH3OH}) derived in this work (blue markers) for 26 high-mass protostars as a function of luminosity. Literature gas-phase values are shown for comparison (red markers), together with ice values in protostars (yellow and black markers), dark clouds (gray dashed line), and comets (purple dash-dot line). The references can be found in Appendix \ref{appendix:lit_ratios}. Upper and lower limits towards protostars are denoted by downward- and upward-facing triangles, respectively. The range of values for dark clouds and comets are shown by their respective shadowed areas. For gas-phase ratios, filled markers correspond to interferometric observations, whereas empty markers denote single dish counterparts. The Pearson correlation coefficient for protostellar ratios in both gas and ices, but excluding single-dish observations and lower limits, is displayed in the upper right corner. IRAS 16293B stands for the source IRAS 16293-2422 B.}
%      \label{fig:NOCS_NCH3OH}
% \end{figure*}

Figure \ref{fig:NOCS_NCH3OH} shows a comparison of the column density ratios of \textit{N}(\ce{OCS})/\textit{N}(\ce{CH3OH}) for various objects. It includes gas-phase observations in low- and high-mass sources as well as ice observations in MYSOs, in dark clouds, and in comets (see Appendix \ref{appendix:lit_ratios} for a list of references). The ratios derived in this work are shown in blue. In general, no trend in ratio versus luminosity is observed. Indeed, the data points for protostars (including both ice and gas phases) result in a Pearson correlation coefficient ($r$) of only -0.11. Spearman correlation tests are also performed and yield similar results to Pearson's for all cases explored in this work. Figure \ref{fig:NOCS_NCH3OH} also contain single-dish observations, but since these probe a much larger scale compared to interferometric counterparts, they are not included in the analysis. Furthermore, \cite{Kushwahaa2023} note that their observations could be subject to beam dilution, which could affect the gas-phase ratios for most of the low-mass sources shown in Figure \ref{fig:NOCS_NCH3OH} (with the exception of IRAS 16293-2422 B). Nonetheless, they can still provide information on general trends in abundances. The fairly constant abundance ratios observed throughout all sources indicate that \ce{OCS} must be formed under similar conditions irrespective of the mass (or luminosity) of the protostar. Considering the drastically different physical conditions experienced by MYSOs and LYSOs during their evolution, particularly regarding their temperatures and UV fields, this lack of correlation with respect to luminosity points at an early formation of the bulk of \ce{OCS} under cold and dense conditions, prior to the onset of star-formation. More data points for low-mass sources would be useful to further constrain any potential trend obscured by the scatter in the ratios.

\begin{figure}[htb!]\centering
\includegraphics[scale=0.23]{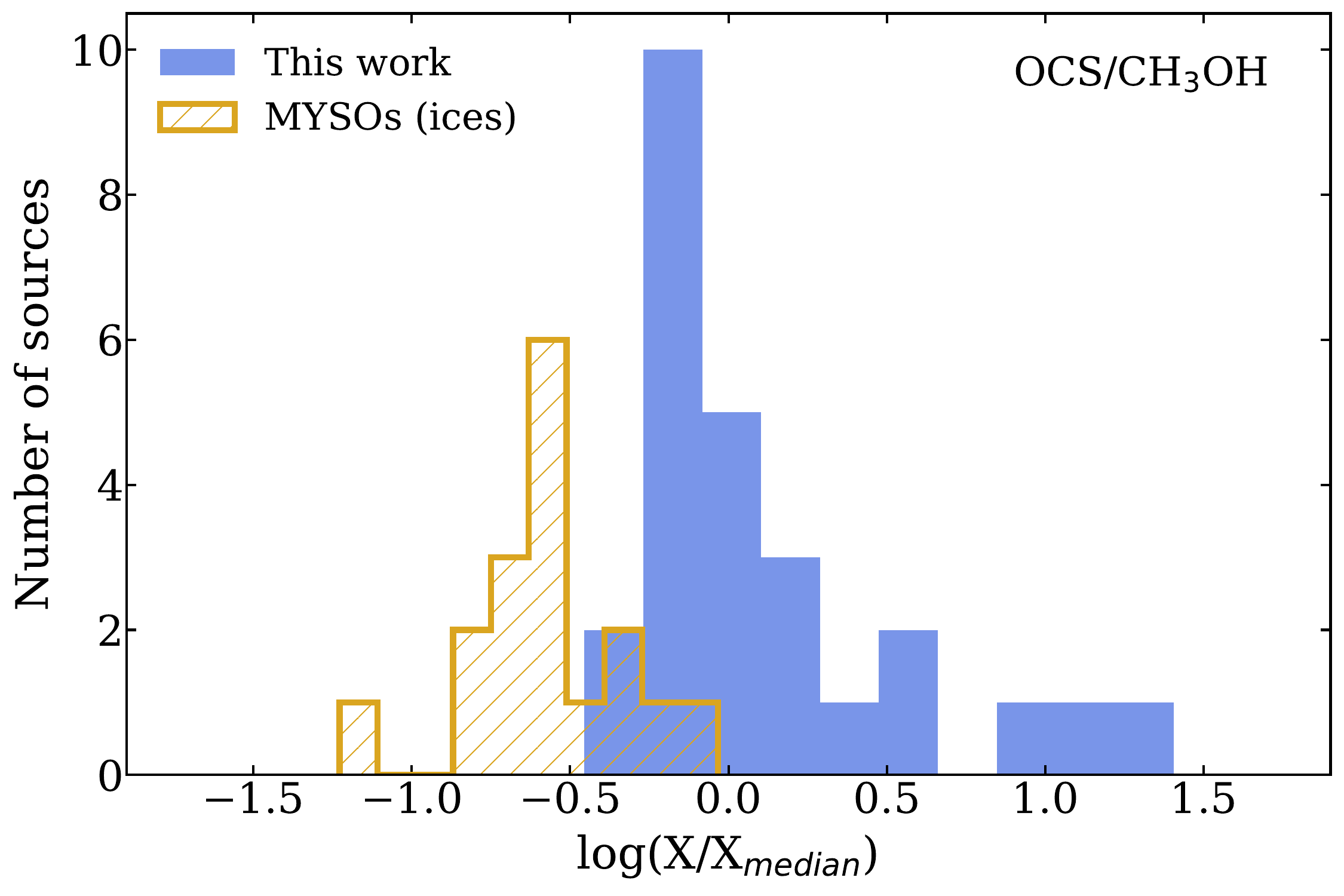}
\caption{Relative abundance distributions of $N$(\ce{OCS})/$N$(\ce{CH3OH}) for the gas-phase observations of massive sources analyzed in this work (blue) compared to ice observations towards 20 MYSOs by \cite{Boogert2022} (yellow). Both histograms are normalized to the weighted gas-phase median value derived from the ALMAGAL dataset for 26 high-mass protostars.}
\label{fig:hist_OCS}
\end{figure}

A considerable number of data points are available for \textit{N}(\ce{OCS})/\textit{N}(\ce{CH3OH}) ratios in protostellar ices \citep{Boogert2022}, which enables a statistically significant analysis. Abundance distribution histograms are an instructive approach to compare different, large datasets (see, e.g., \citealt{Oberg2011}). In Figure \ref{fig:hist_OCS}, the log-transformed \textit{N}(\ce{OCS})/\textit{N}(\ce{CH3OH}) ratios are presented for both ice and gas observations towards MYSOs, and are centered on the weighted median of the ALMAGAL dataset ($\sim$0.033). Ice lower limits affect the median value by $<5\%$ and are thus not included in the analysis. Overall, the ice ratios in MYSOs are slightly lower than the gas-phase counterparts, with a difference of a factor of $\sim$3 between their weighted medians. This discrepancy is remarkably small, and could be due to the uncertainties in the spectral analysis (especially considering the approximation of a fixed $T_{\text{ex}}$). Such strikingly similar ratios provide a strong evidence for an icy origin of \ce{OCS}.

Futhermore, both ice and gas datasets have small scatters, which suggests that the column densities of \ce{CH3OH} and \ce{OCS} are subject to similar dependencies. A spread factor $f$, defined by 10 to the power of the weighted 1$\sigma$ standard deviation measured in log10 space, serves as a convenient comparison basis between distinct datasets (see, e.g., \citealt{Nazari2023a}). Table \ref{tab:medians} summarizes the weighted median gas-phase ratios and spread factors derived in this work for $N$(\ce{OCS})/$N$(\ce{CH3OH}), $N$(\ce{SO2})/$N$(\ce{CH3OH}), and $N$(\ce{SO2})/$N$(\ce{OCS}). For the ALMAGAL subset, the $N$(\ce{OCS})/$N$(\ce{CH3OH}) ratios results in $f=2.8$, whereas ice observations in MYSOs have $f=1.9$. These small scatters suggest similar conditions during the formation of the bulk of \ce{OCS} and \ce{CH3OH}, strengthening the conclusion of an icy origin to \ce{OCS} followed by thermal sublimation. It is worth noting that the lower limits in the available ice ratios are either equivalent to or higher than the corresponding weighted median, so a high abundance tail similar to the gas-phase case could be plausible. 

\begin{table}[htb!]
\centering
\caption{Weighted medians of the column density ratios and spread factors derived in this work for gas-phase species in massive sources (not including upper limits).}
\begin{tabular}{lcc}
\toprule\midrule
Ratio                           &   Weighted median &   Spread factor\\   
\midrule
$N$(\ce{OCS})/$N$(\ce{CH3OH})   &   0.033           &   2.8\\
$N$(\ce{SO2})/$N$(\ce{CH3OH})   &   0.044           &   5.8\\
$N$(\ce{SO2})/$N$(\ce{OCS})     &   1.103           &   3.7\\
\bottomrule
    \end{tabular}
    \label{tab:medians}
\end{table}

The small scatter for \ce{OCS} points to an ice environment similar to \ce{CH3OH}. This agrees with the main proposed chemical routes to form OCS, which involve the sulfurization of \ce{CO} ices. Laboratory experiments show that \ce{OCS} ice can be readily formed by the reactions (\citealt{Ferrante2008, Jimenez-Escobar2014, Chen2015, Nguyen2021, Santos_submm}:

\begin{equation}
    \ce{s-CO} + \ce{s-S} \to \ce{s-OCS}
\label{CO+S}
\end{equation}

\begin{equation}
    \ce{s-CO} + \ce{s-HS} \to \ce{s-OCS} + \ce{s-H},
\label{CO+SH}
\end{equation}

\noindent which can either be induced by thermalized \ce{S} and \ce{SH} adsorbed on the ice or as a result of the energetic processing of larger species (e.g., \ce{CO2} and \ce{H2S}). Indeed, \cite{Boogert2022} analyze a large sample of ice observations towards massive protostars and conclude that \ce{OCS} and \ce{CH3OH} column densities are correlated, pointing to an \ce{OCS} formation concomitant with \ce{CH3OH} during the dense pre-stellar core stage, with both sharing CO as a common precursor.

Comparisons with dark clouds prior to star formation and comets are also relevant to constrain the evolution of sulfur species in both solid and gas phases. Observed $N$(\ce{OCS})/$N$(\ce{CH3OH}) ice ratios in dark clouds agree strikingly well with the protostellar observations towards both ice and gas. This is in full support of the hypothesis that \ce{OCS} is formed in the ices within pre-stellar cores. Caution should be taken when comparing these values, however, since only two data points are available for ices in prestellar cores so far \citep{McClure2023}. Nonetheless, it can still provide the basis for an interesting preliminary comparison. Cometary ratios, in turn, are marginally higher (by a factor of $\sim$4) than the weighted median for ALMAGAL, although their spread encompasses both gas and ice observations. This small discrepancy could be due to additional processing of \ce{OCS} during the protostellar disk phase (as was also suggested by \citealt{Boogert2022}), or as a result of selective \ce{CH3OH} destruction before incorporation into comets. The latter has been previously suggested by \cite{Oberg2011} as one explanation to the depletion of cometary \ce{CH3OH}, \ce{CH4}, and \ce{CO} ices relative to \ce{H2O} compared to protostars. However, given that \ce{OCS} and \ce{CH3OH} are mixed, it seems unlikely that a destruction mechanism could affect one but not the other. 
%%%%%%%%%%%%%%%%%%%%%%%%%%%%%%%%%%%%%%%%%%%%%%%%%%%%%%%%%%%%%%%%%%%%%%%%%%%%%%%%%%%%%%%%%%%%%%%%%%%%%%%%%%%%%%%%%%%%%%%%%%
\subsection{\textit{N}(\ce{SO2})/\textit{N}(\ce{CH3OH})}\label{subsec:NSO2_NCH3OH}
\begin{figure*}[htb!]\centering
\includegraphics[scale=0.42]{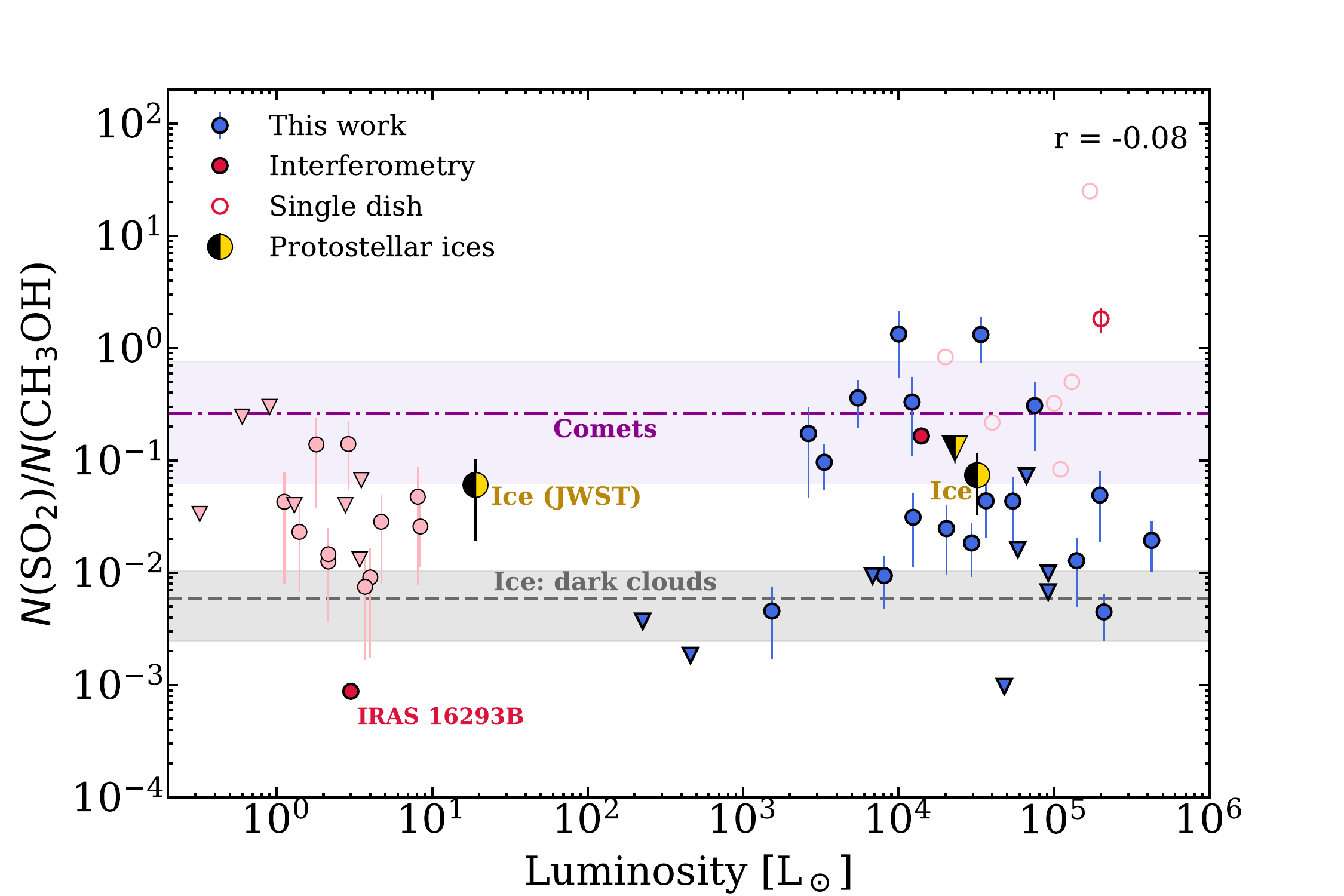}
\caption{Same as Figure \ref{fig:NOCS_NCH3OH}, but for $N$(\ce{SO2})/$N$(\ce{CH3OH}). Literature ratios with $N$(\ce{SO2}) derived from the main isotopologue are signaled by light red markers to differentiate from values derived from $\ce{^{34}SO2}$, which are shown in darker red. The data point for the low-mass protostellar ice ratio corresponds to the JWST observations towards IRAS 2A \citep{Rocha2024}, and the high-mass ice ratios correspond to airborne (KAO) and ground-base (IRTF) observations towards W33A and NGC 7538 IRS 9 (upper limit) \citep{Boogert1997, Boogert2022}.}
\label{fig:NSO2_NCH3OH}
\end{figure*}

%  \begin{figure*}
% \sidecaption
%   \includegraphics[width=12cm]{Figures/NSO2_NCH3OH.pdf}
%      \caption{Same as Figure \ref{fig:NOCS_NCH3OH}, but for $N$(\ce{SO2})/$N$(\ce{CH3OH}). Literature ratios with $N$(\ce{SO2}) derived from the main isotopologue are signaled by light red markers to differentiate from values derived from $\ce{^{34}SO2}$, which are shown in darker red. The data point for the low-mass protostellar ice ratio corresponds to the JWST observations towards IRAS 2A \citep{Rocha2024}, and the high-mass ice ratios correspond to airborne (KAO) and ground-base (IRTF) observations towards W33A and NGC 7538 IRS 9 (upper limit) \citep{Boogert1997, Boogert2022}.}
%      \label{fig:NSO2_NCH3OH}
% \end{figure*}

Figure \ref{fig:NSO2_NCH3OH} shows a comparison of the column density ratios of \textit{N}(\ce{SO2})/\textit{N}(\ce{CH3OH}). Similarly to \ce{OCS}, no correlation is observed for the column density ratios in protostars as a function of luminosity (Pearson $r=-0.08$), in support of the hypothesis of an early formation during the pre-stellar core stage. One caveat to this conclusion is that the gaseous \ce{SO2} column densities in LYSOs are derived from its major isotopologue for most sources (light-red points in Figure \ref{fig:NSO2_NCH3OH}), which could contain appreciable contamination from outflow emission and could be optically thick. In fact, \cite{delaVillarmois2023} estimate that $\sim$40$\%$ of the \ce{SO2} emission in their study is extended. The only exception among low-mass sources is the data point corresponding to IRAS 16293–2422 B, for which \ce{SO2} column densities are derived from \ce{^{34}SO2} (dark-red point in Figure \ref{fig:NSO2_NCH3OH}). Further works focused on compact, optically thin \ce{SO2} emission (as traced by, e.g., \ce{^{34}SO2}) in low-mass sources are warranted to further evaluate this hypothesis.

In contrast to \ce{OCS}, ice ratios of \textit{N}(\ce{SO2})/\textit{N}(\ce{CH3OH}) in protostars are still scarce. However, the available values so far for both high-mass and low-mass sources taken from \cite{Boogert1997, Boogert2022} and \cite{Rocha2024} agree fairly well with the median gas-phase ratios derived in this study. The weighted median for our subset of MYSOs (excluding upper limits) is $\sim$ 0.044, which is consistent with both ice ratios of 0.06$\pm$0.04 and 0.07$\pm$0.04 in a low- and high-mass source, respectively. This agrees with the hypothesis that \ce{SO2} is primarily formed in ices prior to the onset of star formation, and later sublimates upon thermal heating by the protostar.

The scatter in the ALMAGAL values is considerably larger than that for \ce{OCS}, which suggests that the column densities of \ce{CH3OH} and \ce{SO2} are subject to different dependencies. For the ALMAGAL subset, \textit{N}(\ce{SO2})/\textit{N}(\ce{CH3OH}) ratios result in $f=5.8$. For comparison, oxygen and nitrogen-bearing COMs typically have \textit{N}(X)/\textit{N}(\ce{CH3OH}) spread factors of $f\lesssim$3.5 \citep{Nazari2022, Chen2023}. There are likely two reasons behind this large scatter for \ce{SO2}: one related to both \ce{CH3OH} and \ce{SO2} ice environments, and another related to post-desorption processing of \ce{SO2}. 

Compared to methanol, the body of knowledge on interstellar \ce{SO2} formation in cold environments is still somewhat scarce. Currently, its main proposed formation routes generally involve \ce{SO} and a source of oxygen as reactants \citep{Hartquist1980, Charnley1997, Atkinson2004, Blitz2000, Woods2015, Vidal2018, Laas2019}:

\begin{equation}
    \ce{SO} + \ce{OH} \to \ce{SO2} + \ce{H}
\label{SO+OH}
\end{equation}

\begin{equation}
    \ce{SO} + \ce{O2} \to \ce{SO2} + \ce{O}
\label{SO+O2}
\end{equation}

\begin{equation}
    \ce{SO} + \ce{O} \to \ce{SO2}.
\label{SO+O}
\end{equation}

\noindent In the gas phase, such routes are usually not viable at temperatures below 100 K (e.g., \citealt{vanGelder2021}). One exception might be Reaction \ref{SO+OH}, for which rate constants have been predicted to range between $\sim(2-3)\times10^{-10}$ cm$^3$ s$^{-1}$ for temperatures between $10 - 100$ K \citep{Fuente2019}. Still, observed gaseous abundances of \ce{SO2} in both diffuse and dense clouds are $\sim10^{-6}-10^{-5}$ with respect to \ce{CO} (\citealt{Turner1995, Cernicharo2011}, see also Table 4 in \citealt{Laas2019}), which are three to four orders of magnitude lower than ice abundances. Adsorption of \ce{SO2} from the gas phase therefore cannot account for the ice observations. Rather, solid-phase routes to \ce{SO2} must be considered, for which Reactions \ref{SO+OH}$-$\ref{SO+O} are also good candidates \citep{Smardzewski1978, Moore2007, Ferrante2008, Chen2015, Vidal2017}.

Like \ce{SO2}, gas-phase abundances of \ce{SO} with respect to \ce{CO} are still orders of magnitude smaller than those of \ce{SO2} in the ices \citep{Turner1995, Lique2006, Neufeld2015}. Thus, even if all \ce{SO} adsorbed from the gas phase would be converted to \ce{SO2}, it would still be insufficient to explain \ce{SO2} abundances. Alternatively, \ce{S} atoms adsorbed on the ices could react with \ce{O} or \ce{OH} to form \ce{SO} through:

\begin{equation}
    \ce{s-S} + \ce{s-O} \to \ce{s-SO},
\label{S+O}
\end{equation}

\begin{equation}
    \ce{s-S} + \ce{s-OH} \to \ce{s-SO} + \ce{s-H},
\label{S+OH}
\end{equation}

\noindent which in turn can lead to \ce{SO2} ice via Reactions \ref{SO+OH} $-$ \ref{SO+O}. Atomic \ce{S} can also directly form \ce{SO2} ice by reacting with \ce{O2}:

\begin{equation}
    \ce{s-S} + \ce{s-O2} \to \ce{s-SO2}.
\label{S+O2}
\end{equation}

\noindent In addition to the atomic form, interactions of \ce{HS} radicals with \ce{O} can efficiently produce \ce{SO} ice:

\begin{equation}
    \ce{s-HS} + \ce{s-O} \to \ce{s-SO} + \ce{s-H}.
\label{HS+O}
\end{equation}

\noindent Most of these routes have been probed in the laboratory by a number of experimental works involving \ce{H2S} ices mixed with oxygen-bearing molecules (e.g., \ce{CO2} or \ce{H2O}) and exposed to energetic processing (such as UV photons or protons) to generate the open-shell species \citep{Smardzewski1978, Moore2007, Ferrante2008, Chen2015}. Indeed, \ce{HS} radicals are thought to be formed in ices both via the hydrogenation of adsorbed \ce{S} atoms as well as the destruction of \ce{H2S} molecules. The latter can occur either through energetic processing, as mentioned above, or due to \ce{H}-induced abstraction reactions \citep{Oba2018, Oba2019, Santos2023}. Furthermore, \cite{Laas2019} assert that Reactions \ref{SO+O2} and \ref{SO+O} are hindered in ices due to the high diffusion and binding energies of the reactants. Radical and atom formation through energetic processing could partially circumvent this issue.

Irrespective of the mechanism to originate \ce{SO2}, all routes require an oxygen-rich environment to take place. Such an environment is more likely to occur in the earlier phases of the pre-stellar stage, during which \ce{H2O} ices grow from the hydrogenation of \ce{O}, \ce{O2}, and \ce{O3} (e.g., \citealt{Tielens1982, Miyauchi2008, Ioppolo2008, Lamberts2016}). Therefore, \ce{SO2} should be formed simultaneously with \ce{H2O} during the low-density stage of pre-stellar cores. In contrast, \ce{CH3OH} is mainly formed at a much later stage, when densities are high enough for \ce{CO} to catastrophically freeze-out onto the grains ($A_\text{v} > 9$, $n_H \gtrsim 10^5$ cm$^{−3}$). This discrepancy in the formation timeline of the two species means that they will be subject to different physical conditions and collapse timescales, which may explain the large scatter in the observed ratios. While it is true that ice observations of \ce{SO2} have so far been best described by laboratory measurements of \ce{SO2} in a \ce{CH3OH}-rich environment \citep{Boogert1997, Rocha2024}, the \ce{SO2} ice feature at 7.6 $\mu$m was shown to be highly sensitive to ice mixtures and temperatures \citep{Boogert1997}. Hence, further systematic infrared characterizations of \ce{SO2}, perhaps with a combination of tertiary ice mixtures including \ce{H2O}, would be beneficial to better constrain the chemical environment of this molecule.

A complementary explanation for the scatter in $N$(\ce{SO2})/$N$(\ce{CH3OH}) is the reprocessing of \ce{SO2} in the gas phase. The \ce{SO2} emission probed in this work traces the hot core region surrounding the protostar, where thermal heating has led the volatile ice content to fully sublimate. At the typically warmer temperatures of such environments ($\gtrsim$100 K), the conditions become favorable for Reactions \ref{SO+OH}$-$\ref{SO+O} to take place in the gas phase \citep{Hartquist1980, vanGelder2021}. Given the large variation of source structures and physical conditions associated with massive protostars, the degree to which such reactions occur will likely vary considerably from source to source and are thus expected to result in a wide scatter of \ce{SO2} column densities. Indeed, the models in \cite{Vidal2018} predict that the gaseous \ce{SO2} abundances in protostars are particularly subject to large variations depending on the composition of the parent cloud and the temperature. Furthermore, \cite{vanGelder2021} show that gas-phase \ce{SO2} formation is strongly linked to the local UV radiation field, since the strength of the latter will largely affect the distribution of reactants. These properties are expected to vary considerably from one massive source to another.

In addition to \ce{SO2}, \ce{CH3OH} ices typically show considerable variations in column density from source to source \citep{Oberg2011}, which could be contributing to the scatter seen for $N$(\ce{SO2})/$N$(\ce{CH3OH}). Nonetheless, the larger spread for $N$(\ce{SO2}) compared to other species relative to $N$(\ce{CH3OH}) (e.g., \citealt{Nazari2022, Chen2023} and the $N$(\ce{OCS})/$N$(\ce{CH3OH}) ratios in this work) points to a significant effect directly associated with \ce{SO2}. Differences in the emitting area of \ce{SO2} and \ce{CH3OH} could also play a part in the scatter \citep{Nazari2024}, albeit to a lesser extent since experimental laboratory sublimation temperatures of \ce{SO2} and \ce{CH3OH} are generally quite similar (at $\sim$120 K and $\sim$145 K, respectively; \citealt{Kanuchova2017, Mifsud2023, Carrascosa2023}). Codesorption with \ce{H2O} can potentially raise the sublimation temperature of \ce{SO2}, but not enough to result in an appreciable difference to \ce{CH3OH} (see, e.g., \citealt{Fraser2001} for sublimation temperatures of \ce{H2O}). Overall, both the origin of \ce{SO2} in ices and its fate after sublimation are likely to play a part in the large scatter seen in Figure \ref{fig:NSO2_NCH3OH}.

In dark clouds, the ice ratios of $N$(\ce{SO2})/$N$(\ce{CH3OH}) are generally lower than in the protostellar phase (both for ice and gas observations) by about one order of magnitude. This discrepancy could indicate some additional production of \ce{SO2} during the protostellar phase, conceivably due to the enhanced UV radiation field produced in such environments. Cometary ratios are marginally higher, but still in a reasonably good agreement with protostellar ices, and are larger than the weighted median for gas-phase MYSOs by a factor of $\sim6$. As for the case of \ce{OCS}, causes for this enhancement could be additional processing of \ce{SO2} during the protostellar disk phase, or as a result of selective \ce{CH3OH} destruction before incorporation into comets. Given that \ce{SO2} and \ce{CH3OH} are proposed to inhabit different ice phases, this supposition is not a priori unreasonable. In summary, the ratios shown in Figure \ref{fig:NSO2_NCH3OH} support the hypothesis of a moderate, but not complete inheritance of \ce{SO2} ices from the pre-stellar phase into comets.
%%%%%%%%%%%%%%%%%%%%%%%%%%%%%%%%%%%%%%%%%%%%%%%%%%%%%%%%%%%%%%%%%%%%%%%%%%%%%%%%%%%%%%%%%%%%%%%%%%%%%%%%%%%%%%%%%%%%%%%%%%
\subsection{\ce{SO2} versus \ce{OCS}}\label{subsec:XSO2_XOCS}
\begin{figure}[htb!]\centering
\includegraphics[scale=0.23]{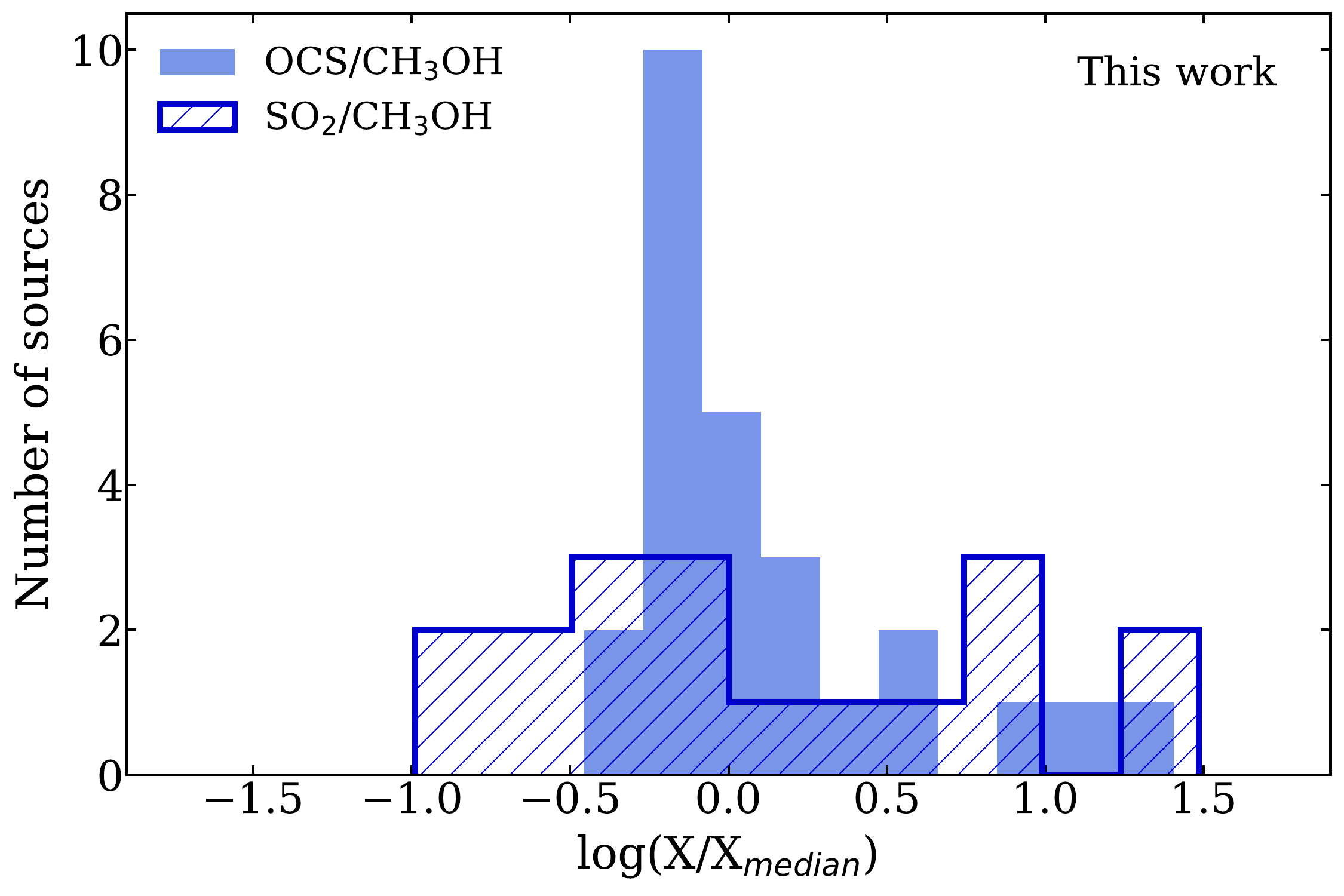}
\caption{Same as Figure \ref{fig:hist_OCS}, but for $N$(\ce{SO2})/$N$(\ce{CH3OH}) (hatched) and $N$(\ce{OCS})/$N$(\ce{CH3OH}) (filled) gas-phase observations from the ALMAGAL dataset. In this case each ratio is normalized to its own weighted median value.}
\label{fig:hist_ALMAGAL}
\end{figure}

The column density ratios for \ce{SO2}/\ce{OCS} as a function of luminosity agree with the conclusions drawn in the previous subsections (see Appendix \ref{appendix:NSO2_NOCS}). Figure \ref{fig:hist_ALMAGAL} presents the distribution histograms of $N$(\ce{SO2}) and $N$(\ce{OCS}) with respect to $N$(\ce{CH3OH}) derived in this work. Each $N$(X)/$N$(\ce{CH3OH}) is normalized to its own weighted median, so that both distributions are centered on 1 (i.e., 0 in log space). This figure clearly shows the distinctive behaviors of \ce{SO2} and \ce{OCS}, emphasizing that the latter is likely much more strongly linked to \ce{CH3OH} than the former. Furthermore, it suggests that reprocessing of \ce{SO2} upon desorption is much more drastic than for \ce{OCS}, in agreement with the models in \cite{Vidal2018}.

\begin{figure}[htb!]\centering
\includegraphics[scale=0.23]{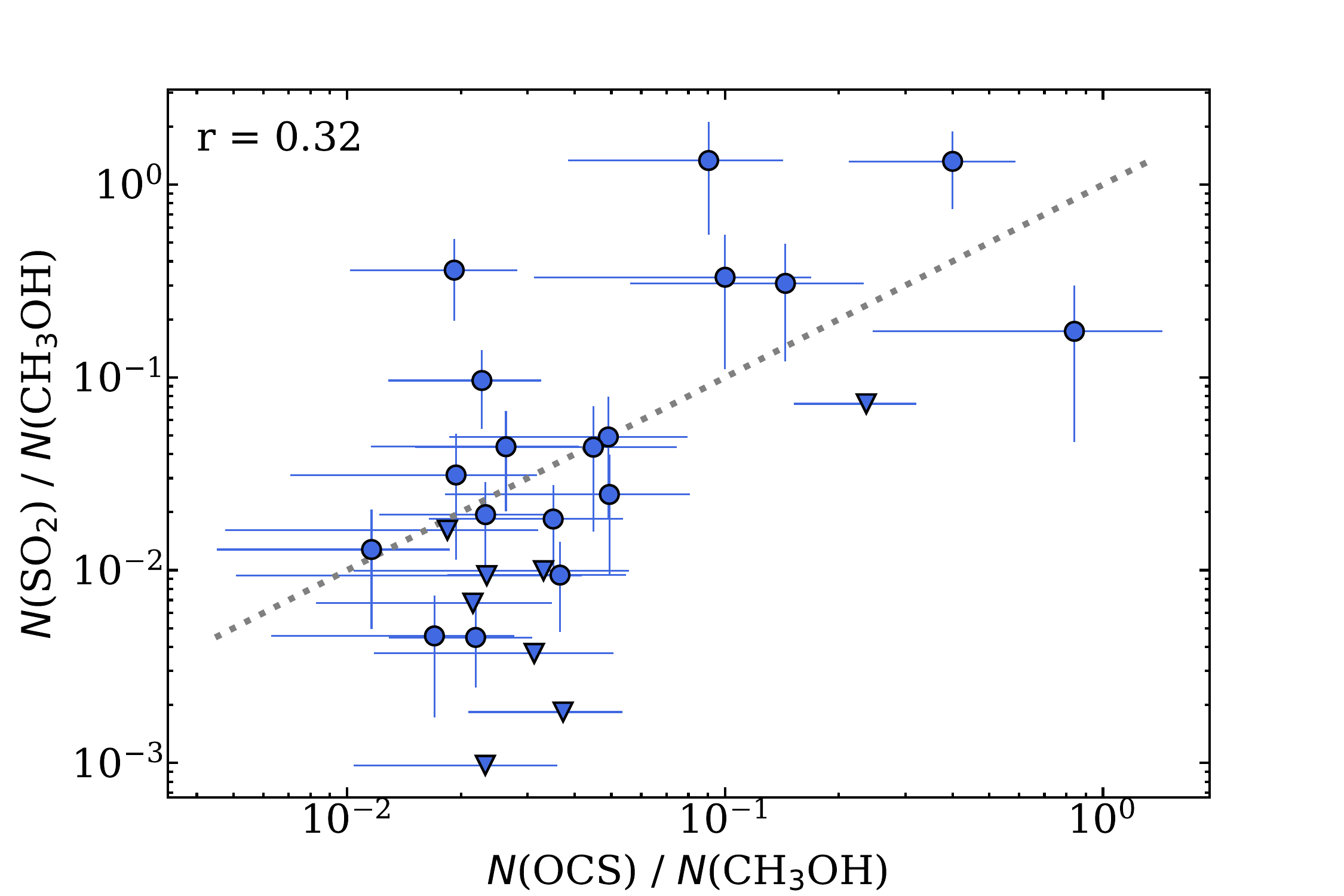}
\caption{Relative abundances $N$(\ce{SO2})/$N$(\ce{CH3OH}) versus $N$(\ce{OCS})/$N$(\ce{CH3OH}) for the ALMAGAL dataset analyzed in this work. Upper limits in the $N$(\ce{SO2})/$N$(\ce{CH3OH}) ratios are denoted by downward facing triangles. The Pearson correlation coefficient (excluding upper limits) is shown in the upper left corner, and the dotted line traces the 1:1 relation.}
\label{fig:XSO2_XOCS}
\end{figure}

Another relevant source of information is to compare the direct correlation between \ce{SO2} and \ce{OCS} abundances with respect to \ce{CH3OH} (Figure \ref{fig:XSO2_XOCS}). Based on the derived Pearson coefficient ($r=0.32$), a weak correlation appears to exist between $N$(\ce{SO2})/$N$(\ce{CH3OH}) and $N$(\ce{OCS})/$N$(\ce{CH3OH}). This is supported by the Spearman's correlation test, which yields a coefficient of $\rho=0.38$ with a p-value of 0.05. The fact that this association is weak is unsurprising, since the bulk of \ce{SO2} and \ce{OCS} probably originate from two different ice environments, which means that their chemistry cannot be strongly linked. Nonetheless, some modest connection seems to be present between the two sulfur-bearing species. This could be due, for instance, to one species providing a source of sulfur that is converted into the other. Indeed, \ce{SO2} ices can be dissociated into \ce{S} atoms upon energetic processing, which could in turn react with \ce{CO} to form \ce{OCS}. Likewise, \ce{OCS} molecules can also yield \ce{S} atoms upon fragmentation, which in turn can react with \ce{H2O} to form \ce{SO2} \citep{Ferrante2008}. Alternatively, this correlation could be the result of a common precursor to both species, such as atomic \ce{S} or \ce{SH} radicals. We emphasize, however, that the correlation is weak and reliant on the high-abundance end of the range of values, and thus should be considered with caution.

%%%%%%%%%%%%%%%%%%%%%%%%%%%%%%%%%%%%%%%%%%%%%%%%%%%%%%%%%%%%%%%%%%%%%%%%%%%%%%%%%%%%%%%%%%%%%%%%%%%%%%%%%%%%%%%%%%%%%%%%%%
%%%%%%%%%%%%%%%%%%%%%%%%%%%%%%%%%%%%%%%%%%%%%%%%%%%% CONCLUSION %%%%%%%%%%%%%%%%%%%%%%%%%%%%%%%%%%%%%%%%%%%%%%%%%%%%%%%%%%
%%%%%%%%%%%%%%%%%%%%%%%%%%%%%%%%%%%%%%%%%%%%%%%%%%%%%%%%%%%%%%%%%%%%%%%%%%%%%%%%%%%%%%%%%%%%%%%%%%%%%%%%%%%%%%%%%%%%%%%%%%
\section{Conclusions}\label{sec:conc}
In this work we analyze the emission of \ce{OCS} and \ce{SO2} towards 26 line-rich MYSOs observed as part of the ALMAGAL survey. We compare their abundances with respect to methanol with other gas-phase observations towards low-mass sources, as well as in interstellar ices and comets. Our main findings are summarized below:
\begin{itemize}

\item The gaseous column density ratios of \ce{OCS}/\ce{CH3OH} and \ce{SO2}/\ce{CH3OH} show no trend with respect to luminosity, pointing at an early onset formation of both sulfur-bearing molecules before star-formation begins. The ratios in protostellar ices are consistent with the weighted medians of the ALMAGAL dataset, suggesting an icy origin to both \ce{OCS} and \ce{SO2} followed by thermal sublimation upon heating from the protostar.

\item A large scatter in relative abundances is observed with ALMAGAL for gaseous \ce{SO2}/\ce{CH3OH} ($f=5.8$), but not for \ce{OCS}/\ce{CH3OH ($f=2.8$)}. Different chemical environments during the formation of \ce{SO2} and \ce{OCS} are proposed as an explanation, with the former being formed during the low-density phase of cold couds, and the latter's formation mostly taking place during the later, high-density pre-stellar stage. \ce{OCS} and \ce{CH3OH} both originate from reactions with \ce{CO} ice. Post desorption processing is also likely to contribute to the spread in $N$(\ce{SO2})/$N$(\ce{CH3OH}).

\item For \ce{OCS}, dark cloud ice values are in remarkably good agreement with both protostellar ice and gas observations. Cometary ratios are also quite similar, at only a factor of $\sim$4 higher. Some extra formation of \ce{OCS} during the protostellar disk phase has been suggested as a root for this difference, as well as selective destruction of \ce{CH3OH}. Nonetheless, all ratios point to a significant inheritance of \ce{OCS} ices throughout the different stages of star formation.

\item The gaseous abundances of \ce{SO2} relative to \ce{CH3OH} derived in this work agree with both dark-cloud and cometary ice ratios. Values in comets are generally slightly higher ($\sim$6$\times$) than in protostars, which in turn are higher than in dark clouds. This could indicate some additional formation of \ce{SO2} during the protostellar and protoplanetary-disk phases, although selective destruction of \ce{CH3OH} could also explain such observations in the latter case.

\item A weak correlation ($r=0.32$) is found between $N(\ce{SO2})/N(\ce{CH3OH})$ and $N(\ce{OCS})/N(\ce{CH3OH})$. While the bulk of these ices is likely formed in different environments during two distinct evolutionary timescales, some interconversion between \ce{SO2} and \ce{OCS} is possible and could lead to a weak association. This could also result from a common precursor among the two species, arguably \ce{S} or \ce{HS}.

\end{itemize}

Overall, our findings suggest that \ce{OCS} and \ce{SO2} differ significantly in both their formation and destruction pathways, but could still potentially share a common history. It should be noted that the dataset studied here is biased in favor of line rich sources and therefore our results might nor represent all massive protostars. Furthermore, it is clear that more observational constraints on interstellar ice column densities of sulfur bearing species, and in particular of \ce{SO2}, are paramount to build a more complete understanding of the origin and fate of sulfur. The \textit{James Webb} Space Telescope offers a unique opportunity for such constraints to be explored in depth.

%%%%%%%%%%%%%%%%%%%%%%%%%%%%%%%%%%%%%%%%%%%%%%%%%%%%%%%%%%%%%%%%%%%%%%%%%%%%%%%%%%%%%%%%%%%%%%%%%%%%%%%%%%%%%%%%%%%%%%%%%%
%%%%%%%%%%%%%%%%%%%%%%%%%%%%%%%%%%%%%%%%%%%%%%%%% ACKNOWLEDGEMENTS %%%%%%%%%%%%%%%%%%%%%%%%%%%%%%%%%%%%%%%%%%%%%%%%%%%%%%%
%%%%%%%%%%%%%%%%%%%%%%%%%%%%%%%%%%%%%%%%%%%%%%%%%%%%%%%%%%%%%%%%%%%%%%%%%%%%%%%%%%%%%%%%%%%%%%%%%%%%%%%%%%%%%%%%%%%%%%%%%%
\begin{acknowledgements}
Astrochemistry at Leiden is supported by funding from the European Research Council (ERC) under the European Union’s Horizon 2020 research and innovation programme (grant agreement No. 101019751 MOLDISK) and the Danish National Research Foundation through the Center of Excellence “InterCat” (Grant agreement no.: DNRF150). 
\end{acknowledgements}

% WARNING
%-------------------------------------------------------------------
% Please note that we have included the references to the file aa.dem in
% order to compile it, but we ask you to:
%
% - use BibTeX with the regular commands:
%   \bibliographystyle{aa} % style aa.bst
%   \bibliography{Yourfile} % your references Yourfile.bib
%
% - join the .bib files when you upload your source files
%-------------------------------------------------------------------

   \bibliographystyle{aa} % style aa.bst
   \bibliography{mybib.bib} % your references Yourfile.bib

%%%%%%%%%%%%%%%%%%%%%%%%%%%%%%%%%%%%%%%%%%%%%%%%%%%%%%%%%%%%%%%%%%%%%%%%%%%%%%%%%%%%%%%%%%%%%%%%%%%%%%%%%%%%%%%%%%%%%%%%%%
%%%%%%%%%%%%%%%%%%%%%%%%%%%%%%%%%%%%%%%%%%%%%%%%%%%% APPENDIX %%%%%%%%%%%%%%%%%%%%%%%%%%%%%%%%%%%%%%%%%%%%%%%%%%%%%%%%%%%%
%%%%%%%%%%%%%%%%%%%%%%%%%%%%%%%%%%%%%%%%%%%%%%%%%%%%%%%%%%%%%%%%%%%%%%%%%%%%%%%%%%%%%%%%%%%%%%%%%%%%%%%%%%%%%%%%%%%%%%%%%%
\begin{appendix}

\section{Example of excluded source}
\label{appendix:bad_source}

\begin{figure}[htb!]\centering
\includegraphics[scale=0.3]{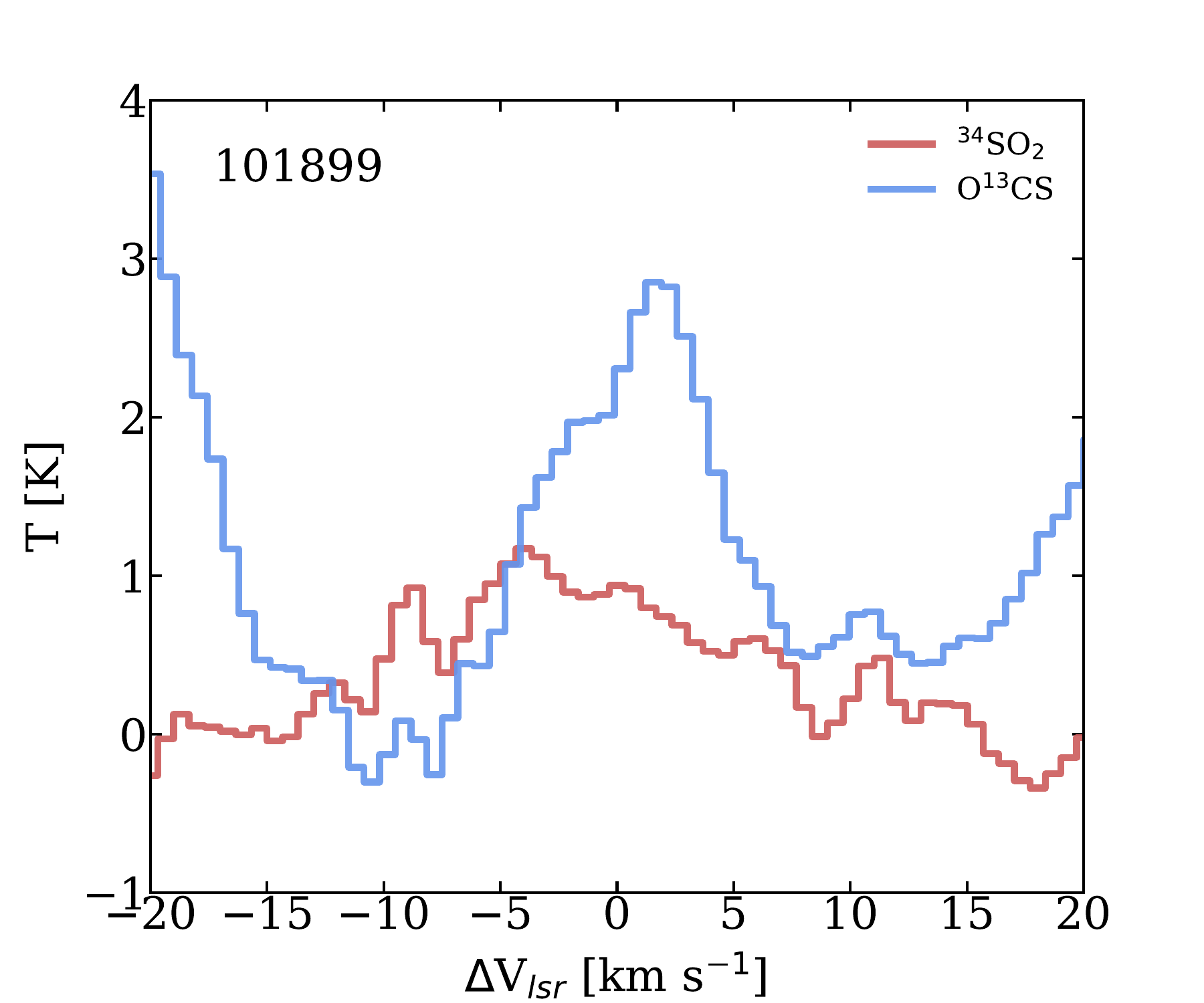}
\caption{Superimposed lines of \ce{^{34}SO2} (red) and \ce{O^{13}CS} (blue) observed towards 101899.}
\label{fig:bad_source}
\end{figure}

Figure \ref{fig:bad_source} contains an example of a source that was excluded from the analysis because of large divergences in the spectral properties between \ce{^{34}SO2} and \ce{O^{13}CS}, in this case potentially due to self absorption of the \ce{^{34}SO2} lines.

%%%%%%%%%%%%%%%%%%%%%%%%%%%%%%%%%%%%%%%%%%%%%%%%%%%%%%%%%%%%%%%%%%%%%%%%%%%%%%%%%%%%%%%%%%%%%%%%%%%%%%%%%%%%%%%%%%%%%%%%%%
\section{Source properties and observational parameters}
\label{appendix:source_stuff}
The observational parameters for each source and their physical properties are listed in Table \ref{tab:source_stuff}.

\begin{table*}[htb!]
\centering
\begin{threeparttable}
\caption{Observational parameters and physical properties of the sources.}
\begin{tabular}{lccccccc}
\toprule\midrule
Source              &   R.A. (J2000) $^a$    &   Dec. (J2000) $^a$    &   Beam                &   Line rms $^b$    &   $d$ $^c$     &   $D_\text{GC}$ $^d$    &   L$_{\text{bol}}$ $^e$\\
                    &   (hh:mm:ss.s)    &   (dd:mm:ss.s)    &   ($\arcsec$)         &   (K)         &   (kpc)   &   (kpc)       &   (L$_{\odot}$)\\
\midrule
126348              &   18:42:51.98     &   -03:59:54.37    &   1.26 $\times$ 1.08 &   0.16         &   4.41    &   4.67        &   6798\\                        
615590              &   09:24:41.96     &   -52:02:08.04    &   0.68 $\times$ 0.59 &   0.50         &   2.70    &   8.31        &   5470\\
644284A             &   10:31:29.78     &   -58:02:19.27    &   0.96 $\times$ 0.80 &   0.32         &   4.75    &   8.20        &   $-$\\
693050              &   12:35:35.05     &   -63:02:31.19    &   1.03 $\times$ 0.94 &   0.19         &   4.31    &   6.89        &   12200\\
705768              &   13:12:36.17     &   -62:33:34.43    &   0.93 $\times$ 0.81 &   0.24         &   6.88    &   6.93        &   91728\\
707948              &   13:16:43.19     &   -62:58:32.83    &   0.94 $\times$ 0.80 &   0.26         &   7.14    &   6.97        &   196800\\
717461A             &   13:43:01.68     &   -62:08:51.42    &   1.34 $\times$ 1.21 &   0.22         &   4.29    &   6.31        &   3323\\
721992              &   13:51:58.27     &   -61:15:41.04    &   0.90 $\times$ 0.79 &   0.37         &   5.38    &   6.16        &   2630\\
724566              &   13:59:30.92     &   -61:48:38.27    &   0.87 $\times$ 0.78 &   0.34         &   4.93    &   6.10        &   226\\
732038              &   14:13:15.05     &   -61:16:53.19    &   0.87 $\times$ 0.78 &   0.40         &   5.64    &   5.93        &   74930\\
744757A             &   14:45:26.35     &   -59:49:15.55    &   1.31 $\times$ 1.27 &   0.15         &   2.51    &   6.45        &   12381\\
767784              &   15:29:19.31     &   -56:31:22.02    &   1.31 $\times$ 1.24 &   0.16         &   4.04    &   5.37        &   139400\\
865468A             &   17:05:10.90     &   -41:29:06.99    &   1.32 $\times$ 1.14 &   0.22         &   3.04    &   5.17        &   47824\\
876288              &   17:11:51.02     &   -39:09:29.18    &   0.92 $\times$ 0.69 &   0.29         &   5.95    &   2.58        &   58410\\
881427C             &   17:20:06.12     &   -38:57:15.84    &   1.31 $\times$ 1.13 &   0.56         &   1.50    &   6.59        &   458\\
G023.3891p00.1851   &   18:33:14.32     &   -08:23:57.82    &   1.43 $\times$ 1.07 &   0.18         &   10.86   &   4.72        &   91560\\
G025.6498p01.0491   &   18:34:20.92     &   -05:59:42.08    &   1.24 $\times$ 1.07 &   0.15         &   12.20   &   6.05        &   424000\\
G305.2017p00.2072A1 &   13:11:10.45     &   -62:34:38.60    &   1.38 $\times$ 1.21 &   0.21         &   4.00    &   6.61        &   20301\\
G314.3197p00.1125   &   14:26:26.25     &   -60:38:31.26    &   1.36 $\times$ 1.23 &   0.21         &   8.25    &   6.33        &   66440\\
G316.6412-00.0867   &   14:44:18.35     &   -59:55:11.28    &   1.32 $\times$ 1.25 &   0.15         &   2.73    &   6.35        &   8080\\
G318.0489p00.0854B  &   14:53:42.64     &   -59:08:53.02    &   1.31 $\times$ 1.27 &   0.15         &   3.18    &   6.07        &   36480\\
G318.9480-00.1969A1 &   15:00:55.28     &   -58:58:52.60    &   1.30 $\times$ 1.27 &   0.15         &   10.40   &   6.83        &   209000\\
G323.7399-00.2617B2 &   15:31:45.45     &   -56:30:49.84    &   1.29 $\times$ 1.25 &   0.15         &   3.20    &   5.79        &   1529\\
G327.1192p00.5103   &   15:47:32.72     &   -53:52:38.60    &   0.90 $\times$ 0.72 &   0.34         &   4.74    &   4.81        &   54270\\
G343.1261-00.0623   &   16:58:17.22     &   -42:52:07.54    &   1.33 $\times$ 1.15 &   0.21         &   2.00    &   6.16        &   33800\\
G345.5043p00.3480   &   17:04:22.89     &   -40:44:23.06    &   1.34 $\times$ 1.15 &   0.20         &   2.00    &   6.13        &   29498\\
\bottomrule
\end{tabular}
\label{tab:source_stuff}
\begin{tablenotes}
\small
\item $^a$ The coordinates correspond to the pixel from which the spectra were extracted.
\item $^b$ Line rms as computed by \cite{Nazari2022}.
\item $^c$ Distance to the source \citep{Lumsden2013, Mege2021}. Typical uncertainties are of $\sim$0.5 kpc.
\item $^d$ Distance to the galactic center assumind the Sun's $D_\text{GC}$ as 8.05 kpc \citep{Honma2015}.
\item $^e$ Bolometric luminosities corrected to the distances $d$ \citep{Lumsden2013, Elia2017}. For regions with multiple cores, the fraction of each individual source was estimated by dividing the total luminosity over all the cores weighted by their peak continuum flux (see also \citealt{vanGelder2022a} and \citealt{Nazari2022}). No estimations for 644284A are available, so a generic L$_{\text{bol}}\sim10^4$ L$_\odot$ is assumed.
\end{tablenotes}
\end{threeparttable}
\end{table*}

%%%%%%%%%%%%%%%%%%%%%%%%%%%%%%%%%%%%%%%%%%%%%%%%%%%%%%%%%%%%%%%%%%%%%%%%%%%%%%%%%%%%%%%%%%%%%%%%%%%%%%%%%%%%%%%%%%%%%%%%%%
\section{List of transitions}
\label{appendix:transitions}
Table \ref{appendix:transitions} lists all the transitions of \ce{^{34}SO2}, \ce{^{33}SO2}, and \ce{O^{13}CS} with $E_{\text{up}}<800$ K and A$_{ij}>10^{-6}$ s$^{-1}$ covered in the data. 

\begin{table*}[htb!]
\centering
\begin{threeparttable}
\caption{Covered transitions.}
\begin{tabular}{ccccc}
\toprule\midrule
Species         &   Transition                                      &   Frequency   &   $E_{\text{up}}$    &   $A_{ij}$\\
                &                                                   &   (MHz)       &   (K)         &   (s$^{-1}$)\\
\midrule
\ce{^{34}SO2}   &   26 8 18 $-$ 27 7 21                             &   217412.915  &   473.565     &   2.09$\times$10$^{-5}$\\ 
\ce{^{34}SO2}   &   31 9 23 $-$ 32 8 24                             &   217902.353  &   646.797     &   2.17$\times$10$^{-5}$\\
\ce{^{34}SO2}   &   11 1 11 $-$ 10 0 10 $^{\star}$              &   219355.009  &   60.075      &   1.11$\times$10$^{-4}$\\
\ce{^{34}SO2}   &   36 5 31 $-$ 35 6 30                             &   220451.866  &   675.942     &   2.61$\times$10$^{-5}$\\
\midrule
\ce{^{33}SO2}   &   6 4 2 4.5 $-$ 7 3 5 5.5                         &   217628.227  &   57.861     &   1.07$\times$10$^{-5}$\\
\ce{^{33}SO2}   &   6 4 2 7.5 $-$ 7 3 5 8.5                         &   217628.440  &   57.861     &   1.04$\times$10$^{-5}$\\
\ce{^{33}SO2}   &   6 4 2 5.5 $-$ 7 3 5 6.5                         &   217628.749  &   57.861     &   1.03$\times$10$^{-5}$\\
\ce{^{33}SO2}   &   6 4 2 6.5 $-$ 7 3 5 7.5                         &   217628.884  &   57.861     &   1.03$\times$10$^{-5}$\\
\ce{^{33}SO2}   &   11 1 11 9.5 $-$ 10 0 10 9.5 $^{\star}$      &   220613.365  &   60.215     &   1.60$\times$10$^{-6}$\\
\ce{^{33}SO2}   &   11 1 11 9.5 $-$ 10 0 10 8.5 $^{\star}$      &   220617.426  &   60.215     &   1.11$\times$10$^{-4}$\\
\ce{^{33}SO2}   &   11 1 11 12.5 $-$ 10 0 10 11.5 $^{\star}$    &   220617.778  &   60.215     &   1.12$\times$10$^{-4}$\\
\ce{^{33}SO2}   &   11 1 11 10.5 $-$ 10 0 10 10.5 $^{\star}$    &   220619.743  &   60.215     &   1.93$\times$10$^{-6}$\\
\ce{^{33}SO2}   &   11 1 11 10.5 $-$ 10 0 10 9.5 $^{\star}$     &   220620.374  &   60.215     &   1.10$\times$10$^{-4}$\\
\ce{^{33}SO2}   &   11 1 11 11.5 $-$ 10 0 10 10.5 $^{\star}$    &   220620.716  &   60.215     &   1.11$\times$10$^{-4}$\\
\ce{^{33}SO2}   &   11 1 11 11.5 $-$ 10 0 10 11.5 $^{\star}$    &   220624.670  &   60.215     &   1.33$\times$10$^{-6}$\\
\midrule
\ce{O^{13}CS}   &   18 $-$ 17 $^{\star}$                        &   218198.998  &   99.49      &   3.01$\times$10$^{-5}$\\
\bottomrule
\end{tabular}
\label{tab:transitions}
\begin{tablenotes}
\small
\item $^{\star}$ Detected transitions. 
\end{tablenotes}
\end{threeparttable}
\end{table*}

%%%%%%%%%%%%%%%%%%%%%%%%%%%%%%%%%%%%%%%%%%%%%%%%%%%%%%%%%%%%%%%%%%%%%%%%%%%%%%%%%%%%%%%%%%%%%%%%%%%%%%%%%%%%%%%%%%%%%%%%%%
\section{Best-fit models for all sources}
\label{appendix:best_models}

\begin{figure*}[htb!]\centering
\includegraphics[scale=0.29]{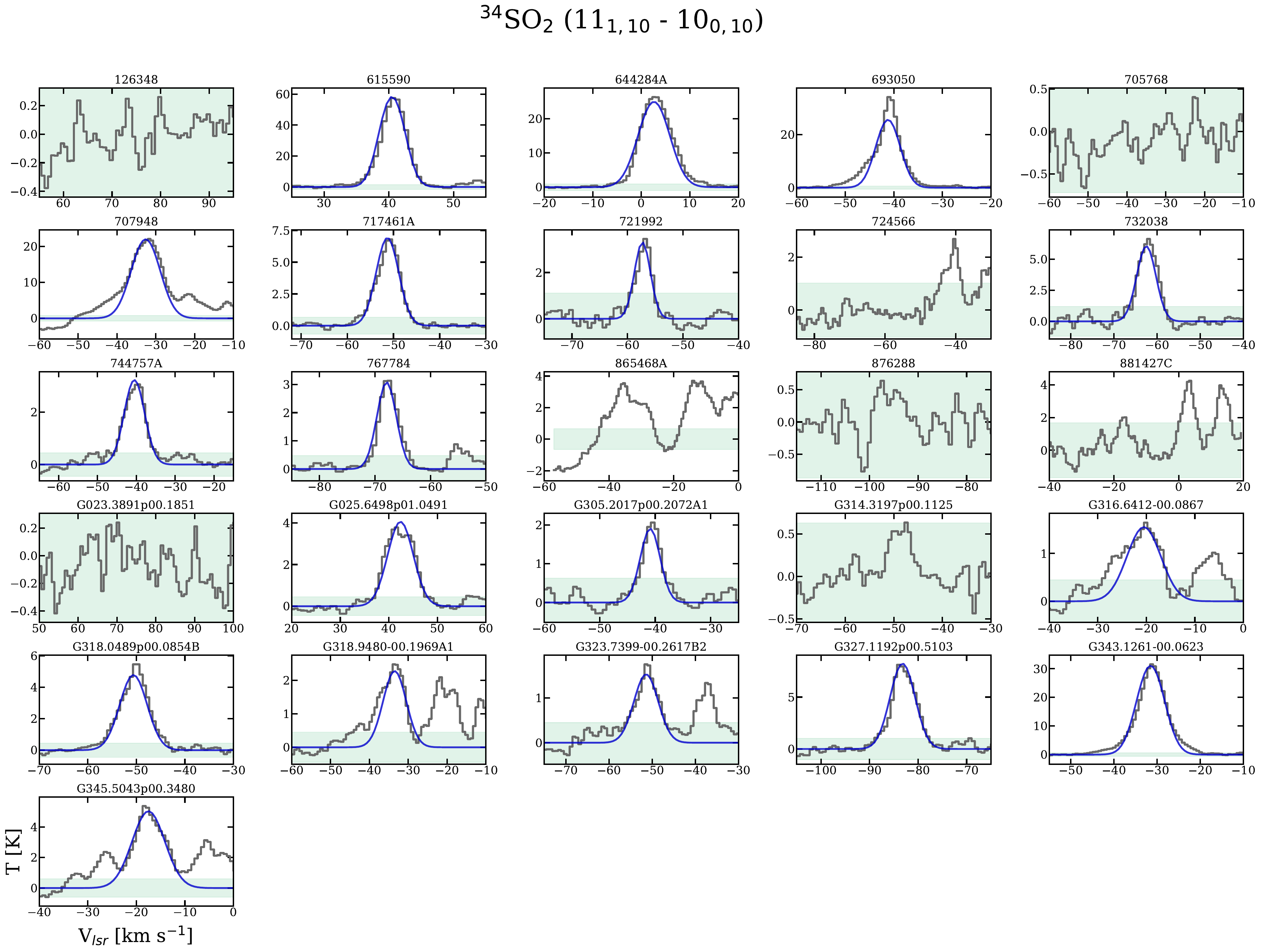}
\caption{Observed spectra towards each source (grey) superimposed by its best-fit model for \ce{^{34}SO2} (blue). The green shadowed area delimits the $3\sigma$ threshold. The name of each source is shown on the top of each panel.}
\label{fig:models_all_34SO2}
\end{figure*}

\begin{figure*}[htb!]\centering
\includegraphics[scale=0.29]{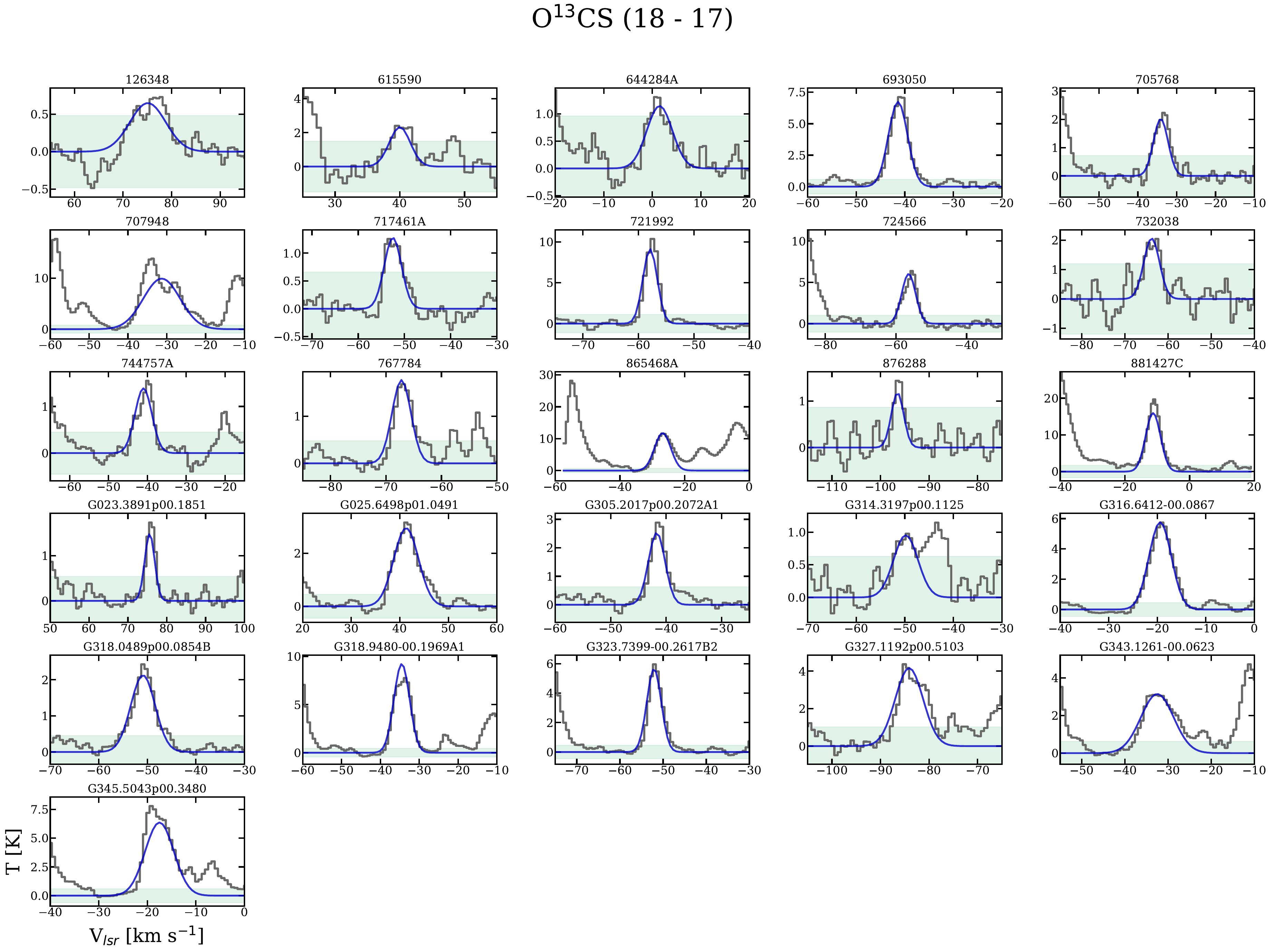}
\caption{Same as Figure \ref{fig:models_all_34SO2}, but for \ce{O^{13}CS}.}
\label{fig:models_all_O13CS}
\end{figure*}

Figures \ref{fig:models_all_34SO2} and \ref{fig:models_all_O13CS} show the best-fit models derived by the grid-fitting approach for each source in the ALMAGAL subset analyzed here. The former contains the results for \ce{^{34}SO2}, and the latter for \ce{O^{13}CS}.

%%%%%%%%%%%%%%%%%%%%%%%%%%%%%%%%%%%%%%%%%%%%%%%%%%%%%%%%%%%%%%%%%%%%%%%%%%%%%%%%%%%%%%%%%%%%%%%%%%%%%%%%%%%%%%%%%%%%%%%%%%
\section{Integrated intensity map of 693050}
\label{appendix:moment0_693050}

\begin{figure}[htb!]\centering
\includegraphics[scale=0.38]{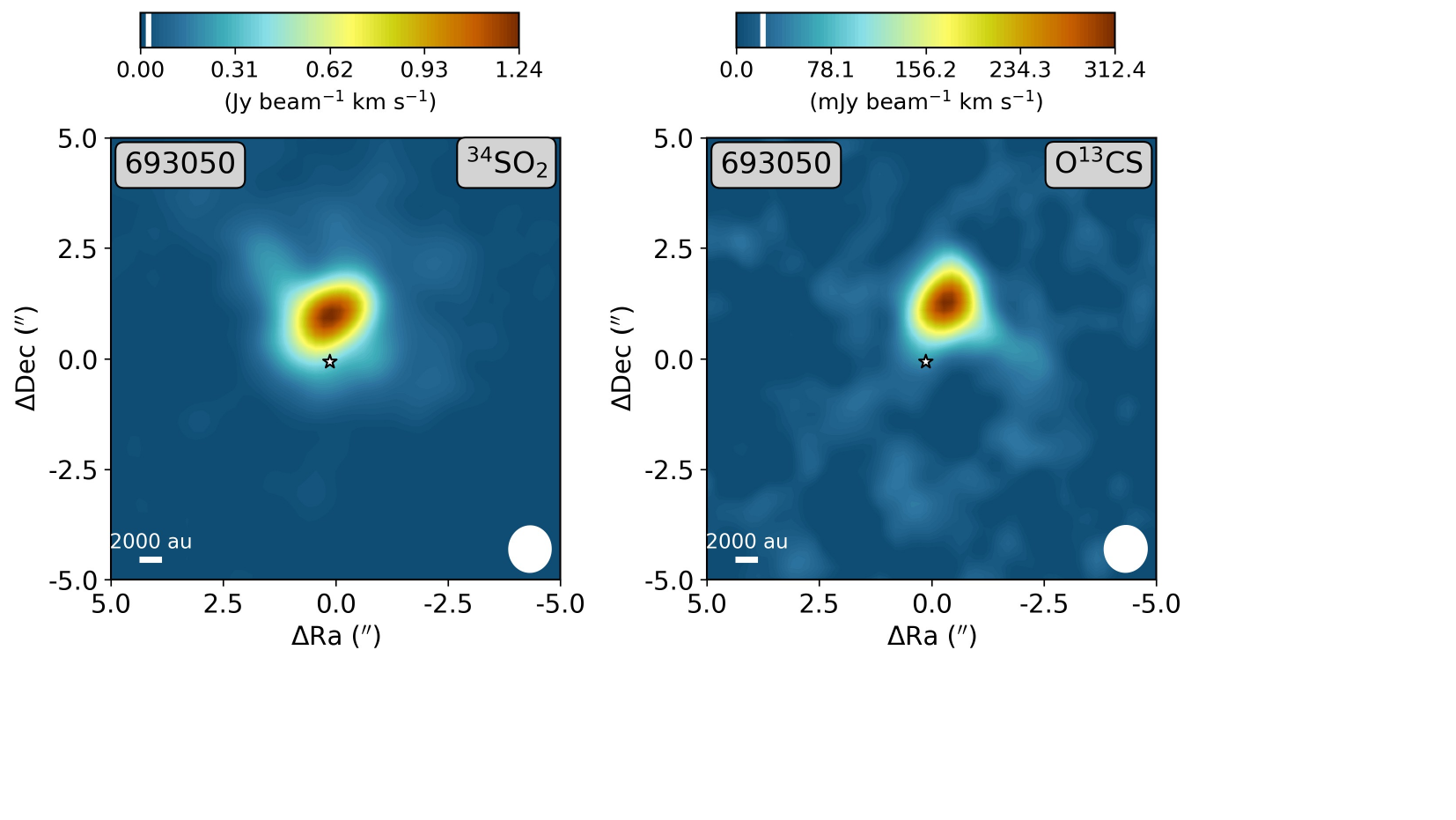}
\caption{Integrated intensity maps of the \ce{^{34}SO2} 11$_{1,11}$ $-$ 10$_{0,10}$ (left) and \ce{O^{13}CS} 18 $-$ 17 (right) lines for 693050. The integration limits are set to [-5, 5] km s$^{-1}$ with respect to the source's $V_{\text{lsr}}$. The white star denotes the source positions derived from the continuum emission and the $3\sigma$ threshold is delimited by the white line in the color bars. The beam size is shown in the lower right corner of each panel, and a scale bar is depicted in the lower left.}
\label{fig:moment0_693050}
\end{figure}

Figure \ref{fig:moment0_693050} shows the integrated intensity maps of source 693050 for \ce{^{34}SO2} 11$_{1,11}$ $-$ 10$_{0,10}$ and \ce{O^{13}CS} 18 $-$ 17. The offset in the emission to the continuum peak is likely due to effects of optically thick dust at these wavelengths.

%%%%%%%%%%%%%%%%%%%%%%%%%%%%%%%%%%%%%%%%%%%%%%%%%%%%%%%%%%%%%%%%%%%%%%%%%%%%%%%%%%%%%%%%%%%%%%%%%%%%%%%%%%%%%%%%%%%%%%%%%%
\section{Best fit parameters of \ce{^{34}SO2} and \ce{O^{13}CS}}
\label{appendix:fits}

Tables \ref{tab:fittings_1} and \ref{tab:fittings_2} list the best-fit parameters for \ce{^{34}SO2} and \ce{O^{13}CS}, respectively, obtained with the grid fitting approach toward all sources. The isotope ratios for each source and corresponding estimated column densities of the main isotopologues are also reported.

\begin{table*}[htb!]
\centering
\begin{threeparttable}
\caption{Fitting results for \ce{^{34}SO2}.}
\begin{tabular}{lccccc}
\toprule\midrule
                    &   \multicolumn{3}{c}{\ce{^{34}SO2}}                                          &   ($^{32}\ce{S}/^{34}\ce{S}$) $^c$      &   \ce{SO2}\\\cline{2-4}
Source              &   $N$ $^a$                           &   FWHM $^b$       &   $V_{\text{lsr}}$ $^b$  &                    &   $N$ $^d$\\
                    &   (cm$^{-2}$)                        &   (km s$^{-1}$)   &   (km s$^{-1}$)   &                                    &   (cm$^{-2}$)\\   
\midrule
126348              &   $<$4.5$\times$10$^{14}$            &   [7.5]           &   76.9            &    20                              &   $<$8.8$\times$10$^{15}$\\                   
615590              &   (4.0$\pm$0.8)$\times$10$^{16}$     &   4.3$\pm$0.4     &   40.5            &    23                              &   (9.0$\pm$2.4)$\times$10$^{17}$\\
644284A             &   (2.5$\pm$0.5)$\times$10$^{16}$     &   7.4$\pm$0.5     &   2.60            &    22                              &   (5.6$\pm$1.5)$\times$10$^{17}$\\
693050              &   (2.0$\pm$0.4)$\times$10$^{16}$     &   5.7$\pm$1.6     &   -41.2           &    22                              &   (4.3$\pm$1.1)$\times$10$^{17}$\\
705768              &   $<$6.0$\times$10$^{14}$            &   [6.7]           &   -34.1           &    22                              &   $<$1.3$\times$10$^{16}$\\
707948              &   (2.5$\pm$0.7)$\times$10$^{16}$     &   8.5$\pm$2.5     &   -32.6           &    22                              &   (5.4$\pm$1.6)$\times$10$^{17}$\\
717461A             &   (5.0$\pm$1.0)$\times$10$^{15}$     &   5.7$\pm$0.5     &   -51.3           &    21                              &   (1.1$\pm$0.3)$\times$10$^{17}$\\
721992              &   (1.3$\pm$0.3)$\times$10$^{15}$     &   3.2$\pm$0.7     &   -57.3           &    21                              &   (2.6$\pm$0.8)$\times$10$^{16}$\\
724566         &   $<$7.0$\times$10$^{14}$            &   [6.0]           &   -57.3           &    21                              &   $<$1.5$\times$10$^{16}$\\
732038              &   (4.0$\pm$0.8)$\times$10$^{15}$     &   5.2$\pm$0.6     &   -62.5           &    21                              &   (8.3$\pm$2.0)$\times$10$^{16}$\\
744757A             &   (2.5$\pm$0.5)$\times$10$^{15}$     &   6.2$\pm$0.7     &   -40.5           &    21                              &   (5.3$\pm$1.3)$\times$10$^{16}$\\
767784              &   (1.6$\pm$0.3)$\times$10$^{15}$     &   4.1$\pm$0.3     &   -67.9           &    20                              &   (3.2$\pm$0.8)$\times$10$^{16}$\\
865468A        &   $<$6.0$\times$10$^{14}$            &   [8.0]           &   -27.3           &    20                              &   $<$1.2$\times$10$^{16}$\\
876288              &   $<$5.0$\times$10$^{14}$            &   4.8$\pm$0.5     &   -96.2           &    18                              &   $<$9.2$\times$10$^{15}$\\
881427C             &   $<$1.1$\times$10$^{15}$            &   [5.5]           &   -10.4           &    18                              &   $<$2.0$\times$10$^{16}$\\
G023.3891p00.1851   &   $<$2.5$\times$10$^{14}$            &   [4.0]           &   75.5            &    21                              &   $<$5.3$\times$10$^{15}$\\
G025.6498p01.0491   &   (3.2$\pm$0.6)$\times$10$^{15}$     &   6.2$\pm$0.9     &   42.4            &    21                              &   (6.6$\pm$1.6)$\times$10$^{16}$\\
G305.2017p00.2072A1 &   (1.0$\pm$0.2)$\times$10$^{15}$     &   4.2$\pm$0.5     &   -40.9           &    21                              &   (2.1$\pm$0.5)$\times$10$^{16}$\\
G314.3197p00.1125   &   $<$3.8$\times$10$^{14}$            &   5.0$\pm$0.5     &   -49.0           &    21                              &   $<$8.0$\times$10$^{15}$\\
G316.6412-00.0867   &   (1.6$\pm$0.3)$\times$10$^{15}$     &   8.2$\pm$0.5     &   -20.5           &    21                              &   (3.3$\pm$0.8)$\times$10$^{16}$\\
G318.0489p00.0854B  &   (4.0$\pm$0.8)$\times$10$^{15}$     &   6.6$\pm$0.4     &   -50.6           &    21                              &   (8.3$\pm$2.1)$\times$10$^{16}$\\
G318.9480-00.1969A1 &   (2.0$\pm$0.5)$\times$10$^{15}$     &   7.0$\pm$4.0     &   -33.5           &    22                              &   (4.3$\pm$1.3)$\times$10$^{16}$\\
G323.7399-00.2617B2 &   (1.3$\pm$0.3)$\times$10$^{15}$     &   6.6$\pm$0.5     &   -51.4           &    21                              &   (2.6$\pm$0.6)$\times$10$^{16}$\\
G327.1192p00.5103   &   (6.3$\pm$1.3)$\times$10$^{15}$     &   6.0$\pm$0.7     &   -83.2           &    20                              &   (1.3$\pm$0.3)$\times$10$^{17}$\\
G343.1261-00.0623   &   (3.2$\pm$0.6)$\times$10$^{15}$     &   7.3$\pm$0.5     &   -31.5           &    21                              &   (6.6$\pm$1.6)$\times$10$^{17}$\\
G345.5043p00.3480   &   (5.0$\pm$1.0)$\times$10$^{15}$     &   7.9$\pm$0.6     &   -17.5           &    21                              &   (1.1$\pm$0.3)$\times$10$^{17}$\\
\midrule\bottomrule
\end{tabular}
\label{tab:fittings_1}
\begin{tablenotes}
\small
\item $^a$ Column densities are calculated using the grid fitting method. $3\sigma$ upper limits are derived by manual fitting.
\item $^b$ Square brackets denote FWHM values kept fixed during the fitting. For the sources where no \ce{^{34}SO2} emission is detected, their FWHM and $V_{\text{lsr}}$ are fixed to the values reported for \ce{CH3^{13}CN} towards the same source \citep{Nazari2022}. For the rest of the sources the velocities are fixed to the best-fitted values derived by the manual approach.
\item $^c$ Values are rounded to zero decimals.
\item $^d$ Column densities of the main isotopologues are derived from the fitting results of the rare isotopologues and considering the isotope ratios.
\end{tablenotes}
\end{threeparttable}
\end{table*}

\begin{table*}[htb!]
\centering
\caption{Fitting results for \ce{O^{13}CS}.}
\begin{tabular}{lccccc}
\toprule\midrule
                    &   \multicolumn{3}{c}{\ce{O^{13}CS}}                                           &   ($^{12}\ce{C}/^{13}\ce{C}$) &   \ce{OCS}\\\cline{2-4}
Source              &   $N$                                 &   FWHM            &   $V_{\text{lsr}}$       &                               &   $N$\\
                    &   (cm$^{-2}$)                         &   (km s$^{-1}$)   &   (km s$^{-1}$)   &                               &   (cm$^{-2}$)\\   
\midrule
126348              &   (5.0$\pm$1.3)$\times$10$^{14}$    &   8.7$\pm$2.3         &   75.1           &  43                          &   (2.2$\pm$0.6)$\times$10$^{16}$\\                   
615590              &   (7.9$\pm$2.1)$\times$10$^{14}$    &   3.8$\pm$1.6         &   40.0           &  60                          &   (4.8$\pm$1.4)$\times$10$^{16}$\\
644284A             &   (6.3$\pm$1.3)$\times$10$^{14}$    &   6.2$\pm$1.7         &   1.50           &  60                          &   (3.8$\pm$0.9)$\times$10$^{16}$\\
693050              &   (2.5$\pm$0.7)$\times$10$^{15}$    &   4.8$\pm$0.5         &   -41.4          &  54                          &   (1.3$\pm$0.4)$\times$10$^{17}$\\
705768              &   (7.9$\pm$2.1)$\times$10$^{14}$    &   4.4$\pm$0.9         &   -34.1          &  54                          &   (4.3$\pm$1.3)$\times$10$^{16}$\\
707948              &   (1.0$\pm$0.3)$\times$10$^{16}$    &   11.0$\pm$2.0        &   -31.3          &  54                          &   (5.4$\pm$1.6)$\times$10$^{17}$\\
717461A             &   (5.0$\pm$1.0)$\times$10$^{14}$    &   4.4$\pm$0.9         &   -51.5          &  51                          &   (2.5$\pm$0.6)$\times$10$^{16}$\\
721992              &   (2.5$\pm$0.5)$\times$10$^{15}$    &   3.0$\pm$0.1         &   -57.9          &  50                          &   (1.3$\pm$0.3)$\times$10$^{17}$\\
724566              &   (2.5$\pm$0.5)$\times$10$^{15}$    &   4.6$\pm$0.6         &   -56.4          &  50                          &   (1.3$\pm$0.3)$\times$10$^{17}$\\
732038              &   (7.9$\pm$1.6)$\times$10$^{14}$    &   4.3$\pm$1.2         &   -63.8          &  49                          &   (3.9$\pm$1.0)$\times$10$^{16}$\\
744757A             &   (6.3$\pm$1.3)$\times$10$^{14}$    &   5.1$\pm$1.2         &   -41.0          &  52                          &   (3.3$\pm$0.8)$\times$10$^{16}$\\
767784              &   (6.3$\pm$1.3)$\times$10$^{14}$    &   4.0$\pm$0.7         &   -67.2          &  46                          &   (2.9$\pm$0.7)$\times$10$^{16}$\\
865468A             &   (6.3$\pm$1.3)$\times$10$^{15}$    &   5.8$\pm$0.5         &   -26.8          &  45                          &   (2.9$\pm$0.7)$\times$10$^{17}$\\
876288              &   (3.2$\pm$0.8)$\times$10$^{14}$    &   3.0$\pm$0.6         &   -96.5          &  33                          &   (1.1$\pm$0.3)$\times$10$^{16}$\\
881427C             &   (7.9$\pm$1.6)$\times$10$^{15}$    &   5.3$\pm$0.4         &   -11.3          &  52                          &   (4.1$\pm$1.0)$\times$10$^{17}$\\
G023.3891p00.1851   &   (4.0$\pm$0.8)$\times$10$^{14}$    &   3.0$\pm$0.1         &   75.6           &  43                          &   (1.7$\pm$0.5)$\times$10$^{16}$\\
G025.6498p01.0491   &   (1.6$\pm$0.3)$\times$10$^{15}$    &   6.0$\pm$0.9         &   41.3           &  50                          &   (7.9$\pm$1.9)$\times$10$^{16}$\\
G305.2017p00.2072A1 &   (7.9$\pm$2.1)$\times$10$^{14}$    &   3.5$\pm$1.0         &   -41.7          &  52                          &   (4.2$\pm$1.2)$\times$10$^{16}$\\
G314.3197p00.1125   &   (5.0$\pm$1.0)$\times$10$^{14}$    &   5.9$\pm$0.8        &   -49.8          &  51                          &   (2.6$\pm$0.6)$\times$10$^{16}$\\
G316.6412-00.0867   &   (2.5$\pm$0.5)$\times$10$^{15}$    &   5.5$\pm$0.4         &   -19.4          &  51                          &   (1.3$\pm$0.3)$\times$10$^{17}$\\
G318.0489p00.0854B  &   (1.0$\pm$0.3)$\times$10$^{15}$    &   5.7$\pm$1.1         &   -50.9          &  50                          &   (5.0$\pm$1.5)$\times$10$^{16}$\\
G318.9480-00.1969A1 &   (4.0$\pm$0.8)$\times$10$^{15}$    &   4.7$\pm$0.3         &   -34.5          &  53                          &   (2.1$\pm$0.5)$\times$10$^{17}$\\
G323.7399-00.2617B2 &   (2.0$\pm$0.4)$\times$10$^{15}$    &   3.9$\pm$0.5         &   -52.1          &  48                          &   (9.7$\pm$2.4)$\times$10$^{16}$\\
G327.1192p00.5103   &   (2.5$\pm$0.5)$\times$10$^{15}$    &   6.7$\pm$1.7         &   -84.1          &  44                          &   (1.1$\pm$0.3)$\times$10$^{17}$\\
G343.1261-00.0623   &   (2.5$\pm$0.7)$\times$10$^{15}$    &   8.9$\pm$2.1         &   -32.5          &  50                          &   (1.3$\pm$0.4)$\times$10$^{17}$\\
G345.5043p00.3480   &   (4.0$\pm$1.0)$\times$10$^{15}$    &   6.9$\pm$1.9         &   -17.5          &  50                          &   (2.0$\pm$0.6)$\times$10$^{17}$\\
\midrule\bottomrule
    \end{tabular}
    \label{tab:fittings_2}
\end{table*}

%%%%%%%%%%%%%%%%%%%%%%%%%%%%%%%%%%%%%%%%%%%%%%%%%%%%%%%%%%%%%%%%%%%%%%%%%%%%%%%%%%%%%%%%%%%%%%%%%%%%%%%%%%%%%%%%%%%%%%%%%%
\section{FWHM and $V_{\text{lsr}}$ of \ce{^{34}SO2} and \ce{O^{13}CS} compared to \ce{CH3OH}}
\label{appendix:FWHMs_Vlsrs_wrtCH3OH}
\begin{figure}[htb!]\centering
\includegraphics[scale=0.27]{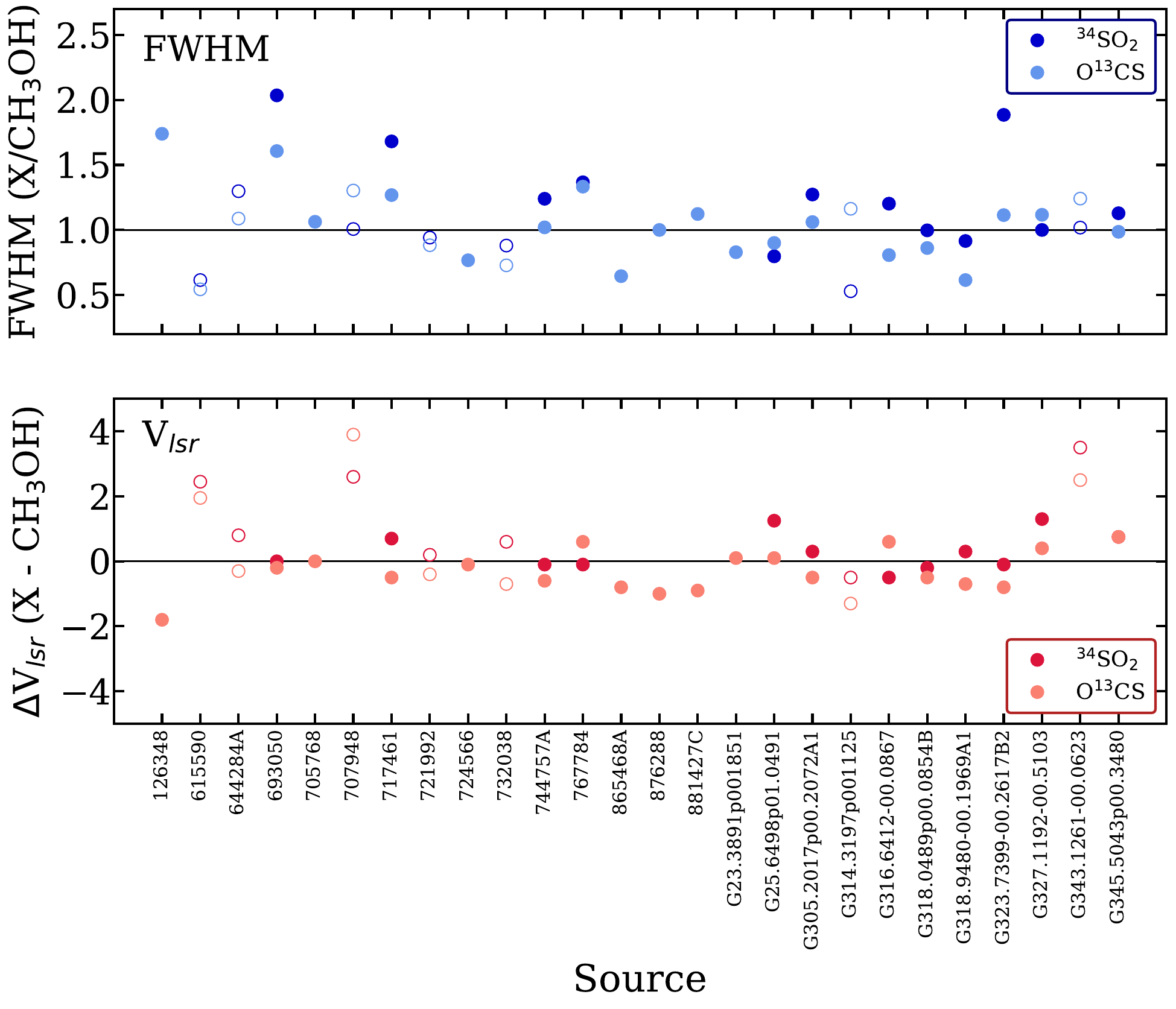}
\caption{FWHM ratios (upper panel) and $V_{\text{lsr}}$ differences (lower panel) with respect to \ce{CH3^{18}OH}. For sources in which \ce{CH3^{18}OH} was not detected by \cite{vanGelder2022}, the main \ce{CH3OH} isotopologue was used (empty markers). Values for \ce{^{34}SO2} are shown in darker colors and for \ce{O^{13}CS} in lighter colors.}
\label{fig:FWHMs_Vlsrs}
\end{figure}

Figure \ref{fig:FWHMs_Vlsrs} shows the ratios of FWHM and differences of $V_{\text{lsr}}$ for \ce{^{34}SO2} and \ce{O^{13}CS} with respect to \ce{CH3OH}.

%%%%%%%%%%%%%%%%%%%%%%%%%%%%%%%%%%%%%%%%%%%%%%%%%%%%%%%%%%%%%%%%%%%%%%%%%%%%%%%%%%%%%%%%%%%%%%%%%%%%%%%%%%%%%%%%%%%%%%%%%%
\section{Best-fit parameters of \ce{^{33}SO2}}
\label{appendix:fitting_33SO2}
Table \ref{tab:fittings_33SO2} lists the best-fit parameters for the \ce{^{33}SO_2} transitions in the sources where it could be constrained.

\begin{table*}[htb!]
\centering
\begin{threeparttable}
\caption{Fitted parameters for \ce{^{33}SO_2}.}
\begin{tabular}{lcccccc}
\toprule\midrule
                    &   \multicolumn{3}{c}{\ce{^{33}SO2}}                                               &   ($^{32}\ce{S}/^{33}\ce{S}$) $^b$   &   \ce{SO2} from \ce{^{33}SO2}     &   \ce{SO2} from \ce{^{34}SO2}\\\cline{2-4}
Source              &   $N$                                 &   FWHM $^a$     &   $V_{\text{lsr}}$ $^a$ &                                               &   $N$                             &   $N$\\
                    &   (cm$^{-2}$)                         &   (km s$^{-1}$) &   (km s$^{-1}$)         &                                               &   (cm$^{-2}$)                     &   (cm$^{-2}$)\\   
\midrule  
615590              &   (1.0$\pm$0.2)$\times$10$^{16}$      &   3.7           &   40.5                  &   93                                          &   (9.3$\pm$0.9)$\times$10$^{17}$  &   (9.0$\pm$2.4)$\times$10$^{17}$\\
644284A             &   (7.0$\pm$1.0)$\times$10$^{15}$      &   6.0           &   2.6                   &   92                                          &   (6.5$\pm$1.4)$\times$10$^{17}$  &   (5.6$\pm$1.5)$\times$10$^{17}$\\
693050              &   (6.0$\pm$1.5)$\times$10$^{15}$      &   3.5           &   -41.2                 &   89                                          &   (5.3$\pm$0.5)$\times$10$^{17}$  &   (4.3$\pm$1.1)$\times$10$^{17}$\\
G343.1261-00.0623   &   (9.5$\pm$1.5)$\times$10$^{15}$      &   7.3           &   -31.5                 &   87                                          &   (8.3$\pm$1.8)$\times$10$^{17}$  &   (6.6$\pm$1.6)$\times$10$^{17}$\\
\bottomrule
\end{tabular}
\label{tab:fittings_33SO2}
\begin{tablenotes}
\small
\item $^a$ Uncertainties in the FWHMs and $V_{\text{lsr}}$ are $\lesssim0.5$ km s$^{-1}$.
\item $^b$ Values are rounded to zero decimals.
\end{tablenotes}
\end{threeparttable}
\end{table*}

%%%%%%%%%%%%%%%%%%%%%%%%%%%%%%%%%%%%%%%%%%%%%%%%%%%%%%%%%%%%%%%%%%%%%%%%%%%%%%%%%%%%%%%%%%%%%%%%%%%%%%%%%%%%%%%%%%%%%%%%%%

\section{Literature ratios}
\label{appendix:lit_ratios}
Table \ref{tab:lit_ratios} lists the literature values for $N$(\ce{SO2})/$N$(\ce{CH3OH}), $N$(OCS)/$N$(\ce{CH3OH}), and $N$(\ce{SO2})/$N$(\ce{OCS}) gathered in this work. 

\begin{table*}[htb!]
\centering
\begin{threeparttable}
\caption{References of literature ratios.}
\begin{tabular}{llll}
\toprule\midrule
Object                   &   $N$(\ce{SO2})/$N$(\ce{CH3OH})          &   $N$(\ce{OCS})/$N$(\ce{CH3OH})       &   $N$(\ce{SO2})/$N$(\ce{OCS})\\  
\midrule
\multicolumn{4}{c}{\underline{Gas phase}}\\
                         &                                          &                                       &\\
MYSOs $^a$               &   (1), (2), (4), (10), (11)              &   (1), (2), (4), (10), (11)           &   (1) $-$ (13)\\
LYSOs $^b$               &   (15), (16), (22), (23)                 &   (15) $-$ (17), (19) $-$ (23)        &   (14) $-$ (16), (18), (22), (23), (24)\\ 
\midrule
\multicolumn{4}{c}{\underline{Interstellar ices}}\\
                         &                                          &                                       &\\
Dark clouds $^c$         &   (25)                                   &   (25)                                &   (25)\\
MYSOs $^d$               &   (26), (27)                             &   (27)                                &   (26), (27)\\
LYSOs $^e$               &   (28), (29)                             &                                       &\\
\midrule
\multicolumn{4}{c}{\underline{Solar system}}\\
                         &                                          &                                       &\\
Comets $^f$              &   (30), (31), (33), (36), (37)           &   (30) $-$ (38)                       &   (30), (31), (33), (36), (37)\\
\bottomrule
\end{tabular}
\label{tab:lit_ratios}
\begin{tablenotes}
\small
\item $^a$ (1) \cite{Hatchell1998}; (2) \cite{vanderTak2000}; (3) \cite{Hofner2001}; (4) \cite{vanderTak2003}; (5) \cite{Osorio2009}; (6) \cite{Herpin2009}; (7) \cite{Cesaroni2010}; (8) \cite{Pillai2011}; (9) \cite{vanderTak2013}; (10) \cite{Gieser2019}; (11) \cite{Fuente2021}; (12) \cite{Mininni2021}; (13) \cite{Fontani2023}.
\item $^b$ (14) \cite{Tafalla2000}; (15) \cite{Jorgensen2018}; (16) \cite{Drozdovskaya2018}; (17) \cite{Sahu2019}; (18) \cite{Agundez2019}; (19) \cite{vanGelder2020}; (20) \cite{Manigand2020}; (21) \cite{Yang2020}; (22) \cite{delaVillarmois2023}; (23) \cite{Kushwahaa2023}; (24) \cite{Esplugues2023}.
\item $^c$ (25) \cite{McClure2023}.
\item $^d$ (26) \cite{Boogert1997}; (27) \cite{Boogert2022}.
\item $^e$ (28) \cite{Tobin2016}; (29) \cite{Rocha2024}.
\item $^f$ (30) \cite{DelloRusso1998}; (31) \cite{Bockelee-Morvan2000}; (32) \cite{Mumma2011}; (33) \cite{LeRoy2015}; (34) \cite{Biver2015}; (35) \cite{DelloRusso2016}; (36) \cite{Calmonte2016}; (37) \cite{Schuhmann2019}; (38) \cite{Saki2020}.
\end{tablenotes}
\end{threeparttable}
\end{table*}
%%%%%%%%%%%%%%%%%%%%%%%%%%%%%%%%%%%%%%%%%%%%%%%%%%%%%%%%%%%%%%%%%%%%%%%%%%%%%%%%%%%%%%%%%%%%%%%%%%%%%%%%%%%%%%%%%%%%%%%%%%
\section{$N$(\ce{SO2})/$N$(\ce{OCS})}
\label{appendix:NSO2_NOCS}
\begin{figure*}
\sidecaption
  \includegraphics[width=12cm]{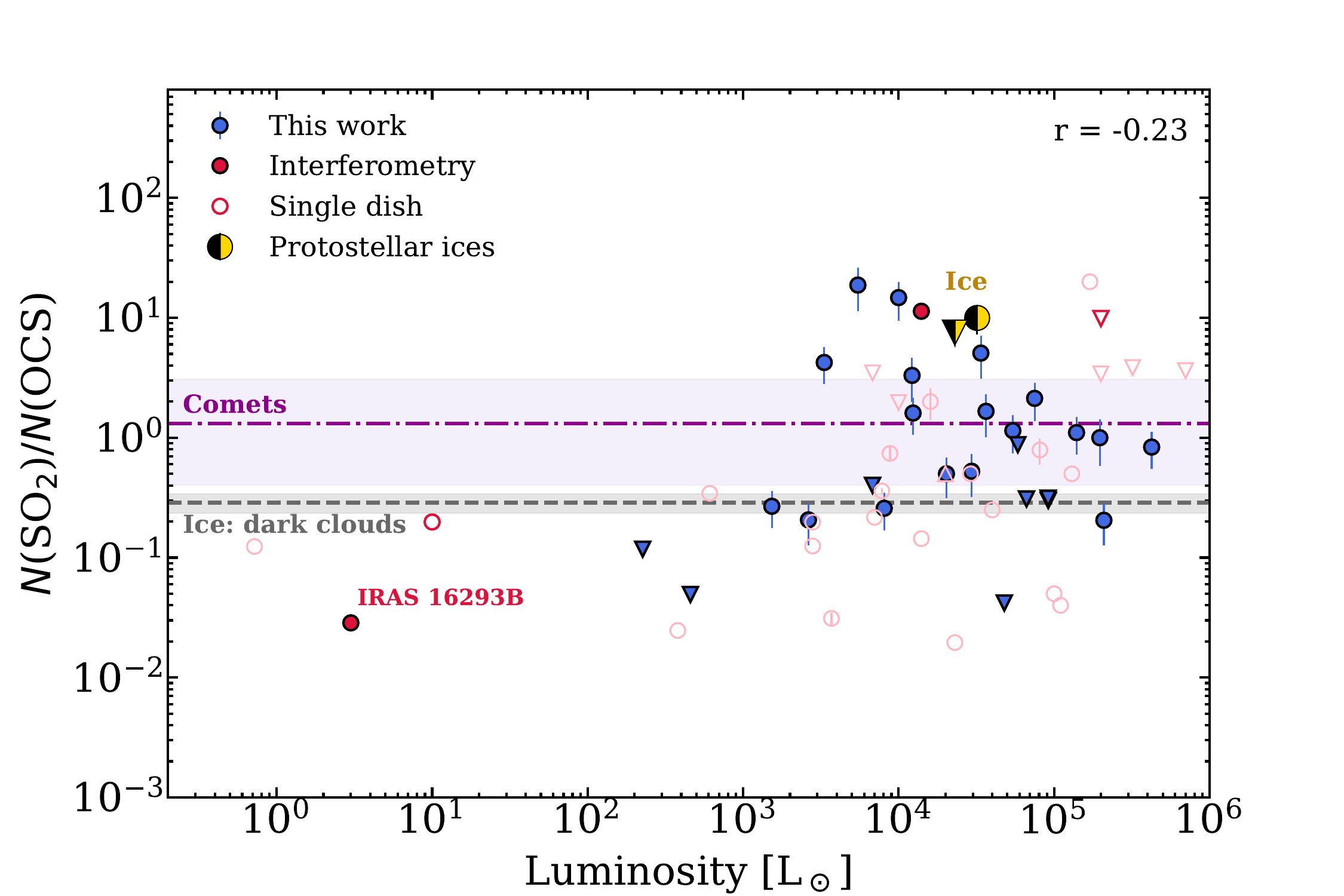}
     \caption{Same as Figure \ref{fig:NOCS_NCH3OH}, but for $N$(\ce{SO2})/$N$(\ce{OCS}). Literature ratios with $N$(\ce{SO2}) derived from the main isotopologue are signaled by light red markers to differentiate from values derived from $\ce{^{34}SO2}$, which are shown in darker red.}
     \label{fig:NSO2_NOCS}
\end{figure*}

The ratios of the derived column densities for \ce{SO2} and \ce{OCS} are shown in Figure \ref{fig:NSO2_NOCS} as a function of luminosity. For comparison, ratios measured in the gas-phase of low-mass sources are also shown, together with ice observations towards comets, dark clouds, as well as both low- and high-mass protostars. 

% \begin{figure*}[htb!]\centering
% \includegraphics[scale=0.42]{Figures/NSO2_NOCS.pdf}
% \caption{Same as Figure \ref{fig:NOCS_NCH3OH}, but for $N$(\ce{SO2})/$N$(\ce{OCS}). Literature ratios with $N$(\ce{SO2}) derived from the main isotopologue are signaled by light red markers to differentiate from values derived from $\ce{^{34}SO2}$, which are shown in darker red.}
% \label{fig:NSO2_NOCS}
% \end{figure*}

\end{appendix}

\end{document}